\documentclass[aps,pre,reprint,floatfix]{revtex4-1}

\usepackage[utf8]{inputenc}
\usepackage[T1]{fontenc}
\usepackage{graphicx}
\usepackage{float}
\usepackage{natbib}
\usepackage{units}
\usepackage{amsmath}
\usepackage{amsfonts}
\usepackage{amssymb}
\usepackage{booktabs}
\usepackage[normalem]{ulem}
\usepackage[usenames,dvipsnames]{color}

\providecommand\nab{\boldsymbol{\nabla}}

\newsavebox{\astrutbox}
\sbox{\astrutbox}{\rule[-5pt]{0pt}{20pt}}

\newcommand{\lap}{\nabla^2}

\renewcommand{\d}{\mathrm{d}}

\begin{document}
\title{Wave turbulence in shallow water models}
\author{P. Clark di Leoni}
\author{P. J. Cobelli}
\author{P. D. Mininni}
\affiliation{Departamento de F\'{\i}sica, Facultad de Ciencias Exactas y 
                  Naturales, Universidad de Buenos Aires and IFIBA, CONICET, 
                  Cuidad Universitaria, Buenos Aires 1428, Argentina}
\date{\today}

\begin{abstract} We study wave turbulence in shallow water flows in numerical
    simulations using two different approximations: the shallow water model, and
    the Boussinesq model with weak dispersion. The equations for both models
    were solved using periodic grids with up to $2048^2$ points. In all simulations,
    the Froude number varies between $0.015$ and $0.05$, while the Reynolds
    number and level of dispersion are varied in a broader range to span different 
    regimes. In all cases, most of the energy in the system remains in the waves, 
    even after integrating the system for very long times. For shallow flows, 
    non-linear waves are non-dispersive and the spectrum of potential energy is 
    compatible with $\sim k^{-2}$ scaling. For deeper (Boussinesq) flows, the 
    non-linear dispersion relation as directly measured from the wave and 
    frequency spectrum (calculated independently) shows signatures of dispersion, 
    and the spectrum of potential energy is compatible with predictions of weak
    turbulence theory, $\sim k^{-4/3}$. In this latter case, the non-linear
    dispersion relation differs from the linear one and has two branches, which
    we explain with a simple qualitative argument. Finally, we study
    probability density functions of the surface height and find that
    in all cases the distributions are asymmetric. The probability
    density function can be approximated by a skewed normal
    distribution as well as by a Tayfun distribution.
\end{abstract}

\maketitle

\section{Introduction} 

Turbulence and non-linear wave interactions in the ocean surface are 
related to important processes in atmospheric sciences and 
oceanography, such as the exchange of energy between the atmosphere 
and the ocean \cite{iafrati_modulational_2013,dasaro_enhanced_2011}. 
This exchange, in turn, plays an important role in the dynamics of the 
planetary and oceanic boundary layers, with consequences on the
transport and mixing of momentum, CO$_2$, and heat 
\cite{ivey_density_2008}. The incorrect modeling of these phenomena 
affects climate evolution predictions
\cite{rose_upper-ocean--atmosphere_2010,cavaleri_wind_2012}. Ocean
surface waves are also of interest in the search of renewable energies
\cite{falnes_review_2007}.

There are several ocean surface models which provide an excellent
framework for studying \textit{weak turbulence theory} 
\cite{hasselmann_non-linear_1962,hasselmann_non-linear_1963,
hasselmann_non-linear_1963-1,zakharov_kolmogorov_1992}. This theory
was developed to describe the out-of-equilibrium behavior of systems
of dispersive and weakly non-linear waves (see, e.g.,
\cite{newell_wave_2011,nazarenko_wave_2011}). Unlike theories of
strong turbulence, for waves and under the assumption of weak
nonlinearities, the equations for two-point correlations can be closed 
and exact solutions with constant flux can be found. Besides this 
assumption, it is also assumed that wave fields are homogeneous, 
and that free waves are uncorrelated.

Weak turbulence theory has been applied to capillary and gravito-capillary waves
\cite{zakharov_kolmogorov_1992}, vibrations on a plate \cite{during_weak_2006},
rotating flows \cite{galtier_weak_2003}, and magnetohydrodynamic waves
\cite{galtier_weak_2000,nazarenko_wave_2011,schekochihin_weak_2012}.  For some
of these systems, the predictions of the theory are compatible with results
obtained from experiments or from numerical simulations. For example, see
Refs.~\cite{deike_decay_2012,falcon_capillary_2009,kolmakov_quasiadiabatic_2004}
for capillary waves, \cite{cobelli_different_2011} for gravitocapillary waves,
\cite{mordant_are_2008,boudaoud_observation_2008,cobelli_space-time_2009} for
vibrations on a plate, and \cite{leerink_multiscale_2012,mininni_energy_2007}
for magnetohydrodynamic waves. Although agreement has been found between theory,
numerical simulations and experiments, there are also discrepancies.  In some of
these cases the compatibility is limited to the spectrum of certain fields (see,
e.g., \cite{mininni_energy_2007}), or to specific configurations used to
generate the excitations. Moreover, for many systems it is also not clear
whether the weak turbulence approximation holds for all times, as the solutions
are not homogeneous in wavenumber space and at sufficiently small scales eddies
may become faster than waves resulting in strong turbulence
\cite{chen_resonant_2005}.

One of the most important applications of weak turbulence is in surface gravity
waves. In oceanography, the Phillips' spectrum \cite{phillips_equilibrium_1958},
derived using dimensional arguments from strong turbulence and considering the
coupling between waves, was long considered to be correct. However, different
observational and experimental data
\cite{toba_local_1973,donelan_directional_1985}, as well as numerical
simulations \cite{badulin_weakly_2007}, suggest that the actual spectrum is
closer to that of weak turbulence. In fact, Phillips himself suggested 
that his spectrum may not be valid in the ocean
\cite{phillips_spectral_1985}. Nonetheless, a scaling compatible with
Phillips' spectrum is still observed in numerical simulations 
\cite{korotkevich_simultaneous_2008} when the forcing is strong. This
suggests that while weak turbulence provides an elegant theoretical
way to study wave turbulence in the ocean, more considerations are 
necessary to appropriately describe the diversity of regimes found 
in these flows \cite{newell_wave_2011}.

Most of the work done in ocean surface waves under the weak turbulence
approximation concerns deep water flows. But the theory can also be applied to the
shallow water case, i.e., for gravity waves whose wavelengths are large compared
to the height of the fluid column at rest (see
\cite{zakharov_statistical_1999,onorato_four-wave_2008}). In this case, the
theory leads to the prediction that the energy spectrum follows a $\sim
k^{-4/3}$ behavior. Behavior compatible with this prediction was found
both experimentally and observationally
\cite{smith_equilibrium_2003,kaihatu_asymptotic_2007,falcon_observation_2011}.
It was also found that an inertial range with a $\sim k^{-2}$ dependency can
develop in the shallower regions of the fluid.  The comparison between different
shallow water models, with different degree of shallowness (and of dispersion)
is therefore of interest, e.g., for the study of waves in basins with
inhomogeneous depth.

In the shallow water regime there are several models that can be
considered to describe the ocean surface. There is the linear theory (see, e.g.,
\cite{landau_fluid_2004}) which can predict the dispersion relation of small
amplitude waves, but which is insufficient to study turbulence. There are also
non-linear theories,  such as the shallow water model \cite{pedlosky_geophysical_1987}
for non-dispersive waves, as well as the Boussinesq model \cite{whitham_linear_1974}
for weakly dispersive waves which is the one used in some of the most recent
works on the subject \cite{onorato_four-wave_2008}. While the former
  non-linear model is valid in the strict shallow water limit, the latter can 
  be used in cases in which the wavelengths are closer to (albeit
  still larger than) the depth of the basin. 

In the present work, we study turbulent solutions 
of the shallow water model and of the Boussinesq equations
using direct numerical simulations.  Previous numerical studies considered the
Hamiltonian equations for a potential flow with a truncated non-linear
    term, or the kinetic equations resulting from weak turbulence theory at
moderate spatial resolution
\cite{dyachenko_weak_2004,korotkevich_simultaneous_2008} 
(with the notable exception of \cite{yokoyama_statistics_2004}). 
Here, we solve the primitive equations, without truncating the 
non-linear terms, potentially allowing for the development of vortical 
motions and of strong interactions between waves, and with spatial 
resolutions up to $2048^2$ grid points.

The paper is organized as follows. In section \ref{equations} we introduce both
models, show the assumptions made in order to obtain them, derive their energy
balance equations, and briefly discuss the predictions obtained in the framework
of weak turbulence theory. In section \ref{simulations} we describe the
numerical methods employed and the simulations. Then, in section \ref{results}
we introduce several dimensionless numbers defined to characterize the flows,
and present the numerical analysis and results. We present wavenumber energy
spectra and fluxes, time resolved spectra (as a function of the wavenumber and
frequency), frequency spectra, and probability density functions of the fluid
free surface height. Finally, in section \ref{conclusion} we present the
conclusions. The most important results are: (a) As in previous experimental and
observational studies \cite{smith_equilibrium_2003,kaihatu_asymptotic_2007} we
find now in simulations that different regimes arise depending on the fluid
depth and the degree of nonlinearity of the system. (b) We obtain a power
spectrum of the surface height compatible (within statistical uncertainties)
with $\sim k^{-2}$ in the shallow water (non-dispersive) case, and one
compatible with a $\sim k^{-4/3}$ spectrum as the fluid depth is increased using
the Boussinesq (weakly dispersive) model. The latter spectrum is also compatible
with predictions of weak turbulence theory \cite{onorato_four-wave_2008}. (c)
Dispersion in the Boussinesq model results in more prominent small scale
features and the development of rapidly varying waves. (d) In the weakly
dispersive regime, the non-linear dispersion relation obtained from the
simulations has two branches in a range of wavenumbers, one branch corresponding
to non-dispersive waves, and another corresponding to dispersive waves. We
interpret this as the result of short wavelength waves seeing an effectively
deeper flow resulting from the interaction with waves with very long wavelength.
(e) The probability density function of surface height can be approximated by
both skewed normal and Tayfun distributions. In the latter case, the parameters
of the distribution are compatible with those previously found in observations
and experiments \cite{falcon_observation_2011}.

\section{The shallow water and Boussinesq equations} 
\label{equations}

\begin{figure}
    \centerline{\includegraphics[width=0.48\textwidth]{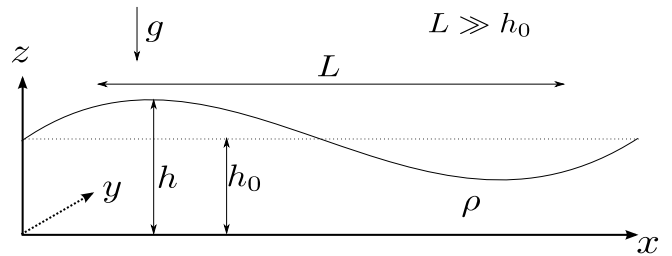}}
    \caption{The shallow water geometry considered in the simulations: $x$ 
        and $y$ are the horizontal coordinates, and $z$ is the vertical 
        coordinate. The surface height is $h$, with $h_0$ the height 
        of the fluid column at rest. The fluid surface is at pressure
        $p_0$. $L$ is a characteristic horizontal length (assumed to
        be much larger than $h_0$). Gravity acts on the $-\hat{z}$ 
        direction and its acceleration has a value of $g$.
        }
    \label{diagram}
\end{figure}

Let us consider a volume of an incompressible fluid with uniform and 
constant (in time) density, with its bottom surface in contact with a 
flat and rigid boundary, and free surface at pressure $p_0$. A sketch
illustrating the configuration is shown in Fig.~\ref{diagram}; $x$ and
$y$ are the horizontal coordinates, $z$ is the vertical one, $h$ is
the height of the fluid column (i.e., the $z$ value at the free surface),
$h_0$ is the height of the fluid column at rest, $L$ is a characteristic
horizontal length, gravity acts on the $-\hat{z}$ direction and its
value is given by $g$. It is assumed that $L \gg h_0$ as we are
interested in shallow flows.

In the inviscid case, both the Euler equation and the incompressibility 
condition hold in the fluid body,
\begin{align}
    \label{euler_basico}
    \frac{\partial \mathbf{v}}{\partial t}
    + (\mathbf{v} \cdot \nab) \mathbf{v}
    &= - \frac{1}{\rho} \nab p - g \hat{z}, \\ 
    \nab \cdot \mathbf{v} &= 0 .
    \label{eq:incompressible}
\end{align}
Under certain assumptions, to be discussed in the following paragraphs,
    the evolution of the free surface can be adequately described by means of a
    vector equation for the two horizontal components of the velocity at the
    interface, plus an equation for the local height of the fluid column.

\subsection{Linear dispersion relation} 

Considering the case of very small amplitude waves, one can linearize 
the system of Eqs.~\eqref{euler_basico} and \eqref{eq:incompressible} 
(see, e.g., \cite{landau_fluid_2004}). The solutions of the resulting 
equations are gravity waves with the following dispersion relation
\begin{equation}
    \label{rel_disp_tot}
    \omega^2 = gk \frac{1 - e^{-2kh_0}}{1 + e^{-2kh_0}} .
\end{equation}

We are interested in the shallow water case, i.e., when $h_0 \ll
L \Rightarrow h_0 k \ll 1 $. In that limit the following
dispersion relation results
\begin{equation}
    \label{rel_disp_sw}
    \omega = \sqrt{g h_0} k = c_0 k ,
\end{equation}
where $c_0 = \sqrt{g h_0}$ is the phase velocity. Note in this case
waves are not dispersive, unlike the general case given by Eq.~\eqref{rel_disp_tot}.

\subsection{Shallow water model} 

It is possible to derive a set of non-linear equations for the surface
height and the horizontal velocity at the surface by using the fact that the
fluid layer is shallow. Considering the characteristic magnitudes of all
quantities in Eq.~\eqref{euler_basico} ($L$, $p_0$, $h_0$,
    $g$, etc.), and using the fact that in a shallow flow $h_0 k \ll 1$ 
with $k=2\pi/L$, one obtains hydrostatic balance in the vertical
direction (for further details, see 
\cite{pedlosky_geophysical_1987}), which results in the pressure profile
\begin{equation}
    p = \rho g (h-z) + p_0 .
\end{equation}

As $h$ is not a function of $z$, neither will be the horizontal pressure 
gradient and the horizontal components of the velocity (as long as
they do not depend initially on $z$). In this way, the horizontal 
components of Eq.~\eqref{euler_basico} can be written as
\begin{equation}
    \label{eqs_horz}
    \frac{\partial \mathbf{u}}{\partial t} = -(\mathbf{u} \cdot \nab) \mathbf{u}
    -g \nab h .
\end{equation}
where $\mathbf{u}(x,y,t) = v_x \hat{x} + v_y \hat{y}$ is the
horizontal velocity, and $\nab$ is now the horizontal gradient.

Using the fact that $v_x$ and $v_y$ are independent of $z$ we can integrate 
the incompressibility condition, obtaining
\begin{equation}
    v_z (x,y,z,t) = -z \left( \frac{\partial v_x}{\partial x} + \frac{\partial
            v_y}{\partial y} \right) + \tilde{v}_z (x,y,t) .
\label{eq:vz}
\end{equation}

Finally, by taking the appropriate boundary conditions and setting 
$z=h(x,y,t)$, Eq.~\eqref{eq:vz} provides an equation for the
evolution of the height of the fluid column, namely
\begin{equation}
    \label{eq_h}
    \frac{\partial h}{\partial t} + \frac{\partial }{\partial x} (h v_x)
    + \frac{\partial }{\partial y}(h v_y) = 0 .
\end{equation}
Note that we do not have to assume irrotationality to derive neither
Eq.~\eqref{eqs_horz} nor Eq.~\eqref{eq_h}.

If we linearize these equations, we find again the dispersion relation given by
Eq.~\eqref{rel_disp_sw}, as expected. In the presence of external forcing
$\mathbf{F}$, and  viscosity $\nu$, the equations can be written as
\begin{align} \label{sw_u} \frac{\partial \mathbf{u}}{\partial t} &=
    -(\mathbf{u} \cdot \nab) \mathbf{u}  -g \nab h + \frac{\nu}{h} \nab \cdot (
    h \nab \mathbf{u}) + \mathbf{F}, \\ \label{sw_h} \frac{\partial h}{\partial
        t} &= -\nab \cdot (h \mathbf{u}) .  \end{align} We will refer to this
set of equations as the shallow water model, or SW model. In these equations the
viscosity $\nu$ was added to the horizontal velocity field $\mathbf{u}$, which
behaves as a compressible flow (i.e., it has non-zero divergence, see
\cite{marche_derivation_2007}). This choice of the viscous term ensures
conservation of the momentum $h \mathbf{u}$, and also that energy dissipation is
always negative, as in Sec.~\ref{energybalancesec}.

\subsection{Boussinesq model} 

As the depth of the fluid increases, dispersion becomes important. There are
several models that introduce weak dispersive effects perturbatively, but many
are built to study waves propagating in only one direction. The Boussinesq
equations for surface waves (see, e.g.,
\cite{whitham_linear_1974,choi_nonlinear_1995}) provide a model to study weakly
dispersive waves propagating in any direction. This model not only
    broadens the range of physical phenomena encompassed by the SW model, but
    adding dispersive effects also makes it more enticing to weak turbulence
    theory, for which dispersion effects are of the utmost importance.

Let us take a look at the first terms in the Taylor
expansion of the dispersion relation in Eq.~\eqref{rel_disp_tot},
\begin{equation}
    \label{taylor_exp}
    \omega^2 = c^2_0 k^2 - \frac{1}{3} c^2_0 h^2_0 k^4 + \ldots ,
\end{equation}
where the first term is the non-dispersive shallow water term. The
idea is to add terms to Eqs.~\eqref{sw_u} and \eqref{sw_h} such that 
the linear dispersion relation of the new system coincides, up to the 
fourth order, with the expansion in Eq.~\eqref{taylor_exp}. This can
be done by adding the term 
$h^2_0 \lap \partial_t \mathbf{u}/3$ to Eq.~\eqref{sw_u},
resulting in the following system,
\begin{align}
    \frac{\partial \mathbf{u}}{\partial t} =& -(\mathbf{u} \cdot \nab) \mathbf{u}  -g
    \nab h + \frac{1}{3} h^2_0 \lap \frac{\partial \mathbf{u}}{\partial t} \nonumber \\
{}& + \frac{\nu}{h} \nab \cdot ( h \nab \mathbf{u}) + \mathbf{F}, \label{bq_u} \\
    \label{bq_h}
    \frac{\partial h}{\partial t} =& -\nab \cdot (h \mathbf{u}) .
\end{align}
We will refer to this system as the Boussinesq model, or BQ model. For
 $\mathbf{F}=0$ and $\nu=0$, the dispersion relation obtained
by linearizing these equations is
\begin{equation}
    \label{rel_disp_bous}
    \omega = \frac{c_0 k }{\sqrt{1 + \frac{h^2_0 k^2}{3}}} ,
\end{equation}
which, up to the fourth order, coincides with Eq.~\eqref{rel_disp_tot}. 

Note that there are other choices for the extra term in
Eq.~\eqref{sw_u} that result in many formulations of the Boussinesq
model, all compatible up to fourth order in a Taylor expansion in terms
of $h_0 k$ \cite{whitham_linear_1974}. The formulation we use here
was employed in previous studies of wave turbulence 
\cite{onorato_four-wave_2008}, and is also easy to solve numerically
using pseudospectral methods by writing Eq.~\eqref{bq_u} as
\begin{equation}
\frac{\partial \mathbf{u'}}{\partial t} = -(\mathbf{u} \cdot \nab) \mathbf{u}  -g
    \nab h + \frac{\nu}{h} \nab \cdot ( h \nab \mathbf{u}) +
    \mathbf{F},
\label{eq:Helmholtz}
\end{equation}
where $\mathbf{u'} = {\cal H}\mathbf{u}$, and where 
${\cal H} = (1-h_0^2 \lap /3)$ is the Helmholtz operator. This
operator can be easily inverted in Fourier space 
\cite{mininni_numerical_2005,mininni_numerical_2005-1}, and the resulting 
equations can be efficiently solved by means of pseudospectral codes. It is
interesting that the same operator appears in Lagrangian-averaged 
models \cite{foias_navier-stokes-alpha_2001}. In these models, and in 
regularized versions of the shallow water equations 
\cite{camassa_integrable_1993}, it introduces dispersion that results
in an accumulation of energy at small scales \cite{graham_highly_2007}.

\subsection{Energy balance} 
\label{energybalancesec}

An exact energy balance can be easily derived for the SW model. The
equation is useful to verify conservation in pseudospectral codes. By 
taking the dot product of Eq.~\eqref{sw_u} and $h \mathbf{u}$, setting
${\bf F} = 0$, and using Eq.~\eqref{sw_h}, we obtain
\begin{equation}
\begin{aligned}
    \frac{\partial }{\partial t} \left( \frac{h u^2}{2}  + g \frac{h^2}{2}
    \right) = &- \nab \cdot \left( \frac{h u^2}{2} \mathbf{u}  + g h^2 \mathbf{u}
    \right) \\
    &+ \nu \mathbf{u} \cdot [ \nab \cdot ( h \nab \mathbf{u}) ] .
\end{aligned}
\end{equation}
Integrating in $x$ and $y$ over an area $A$ and taking periodic
boundary conditions yields
\begin{equation}
    \label{balance}
    \frac{\mathrm{d} E}{\mathrm{d} t} = - 2 \nu Z,
\end{equation}
where
\begin{equation}
E = \frac{1}{A} \iint  \left( \frac{h u^2}{2} + g \frac{h^2}{2} \right)
    \mathrm{d} x \mathrm{d} y
\end{equation}
is the mean total energy, and
\begin{equation}
Z = \frac{1}{A} \iint  \frac{h \lvert \nab \mathbf{u}
    \rvert^2}{2}  \mathrm{d} x \mathrm{d} y
\end{equation}
is a mean pseudo-enstrophy, such that $-2 \nu Z$ is the mean 
energy dissipation rate. As $h$ is always positive, the energy
dissipation is always negative. The total energy is conserved 
when $\nu = 0$.

Now we can define
\begin{equation}
    U = \frac{1}{A} \iint \label{ec_cin} 
    \frac{h u^2}{2}\mathrm{d} x \mathrm{d} y
\end{equation}
as the mean kinetic energy, and
\begin{equation}
    \label{ec_pot}
    V = \frac{1}{A} \iint g \frac{h^2}{2} \mathrm{d} x \mathrm{d} y
\end{equation}
as the mean potential energy, such that the sum of both gives the 
mean total energy $E$.

The dispersive term present in the BQ model changes the balance given by
Eq.~\eqref{balance}. However, since the extra term is of order $(h_0/L)^2$, as
long as we are in a sufficiently shallow flow it will be very small, and
therefore, negligible for the conservation of energy. We verified this is the
case in our numerical simulations.

\subsection{\label{sec:weak}Weak Turbulence prediction} 

We briefly present some results obtained in the framework of weak turbulence
theory for the BQ model (as the derivation is a bit cumbersome, only a
    general outline will be given here; please see
\cite{onorato_four-wave_2008} for details). Weak turbulence is studied in the BQ
model assuming the fluid is inviscid and irrotational, so that the velocity can
be written in terms of a velocity potential. To obtain a statistical description
of the wave field, it is also assumed that it is homogeneous and that the free
modes are uncorrelated.

At first sight, the quadratic
nonlinear terms in Eqs.~\eqref{bq_u} and \eqref{bq_h} indicate modes
interact in triads, with the wave vectors of the three interacting
modes lying over a triangle, and the three frequencies satisfying the
resonant condition (see, e.g., \cite{nazarenko_wave_2011})
\begin{align}
    \mathbf{k} &= \mathbf{p} + \mathbf{q} \\
    \omega(\mathbf{k}) &= \omega(\mathbf{p}) + \omega(\mathbf{q}) .
\end{align}
However, as there are no three wave vectors $\mathbf{k},
\mathbf{p}, \mathbf{q}$ that satisfy these two conditions 
when the dispersion relation is given by Eq.~\eqref{taylor_exp}, three 
wave interactions are forbidden. Thus, only four wave interactions are
present (which do satisfy their corresponding condition).

After a transformation of the fields, it is possible to write an
equation for the evolution of the two-point correlator of the 
transformed fields. 
This is the so-called kinetic equation, and has the 
following form
\begin{gather}
\begin{aligned}
    \frac{\partial N_0}{\partial t} = 4 \pi &\int \lvert T_{0,1,2,3} \rvert^2
    N_0 N_1 N_2 N_3 \\
    & \left( \frac{1}{N_0} + \frac{1}{N_1}- \frac{1}{N_2}-
    \frac{1}{N_3} \right) \\
        & \delta (\mathbf{k}_0 + \mathbf{k}_0 - \mathbf{k}_2 - \mathbf{k}_3) \\
        &\delta (\omega_0 + \omega_0 - \omega_2 - \omega_3) \textrm{d}
        \mathbf{k}_{123},
        \label{kinetic_eq}
\end{aligned}
\end{gather}
where $N_i = N(\mathbf{k}_i)$ is the wave action spectral density
(i.e., the two-point correlator of the wave action, the latter being a quantity proportional to the surface height), the 
deltas express the fact that interactions are between four wave vectors
and their associated frequencies, $T_{0,1,2,3}$ is the coupling
coefficient between the four modes, and 
$\textrm{d} \mathbf{k}_{123} = \textrm{d}
\mathbf{k}_{1} \textrm{d} \mathbf{k}_{2} \textrm{d} \mathbf{k}_{3}$. 
From this equation, dimensional analysis yields the
following expression for the energy spectrum
\begin{equation}
    \label{predicc_wt}
    E(k) \sim k^{-4/3} .
\end{equation}
From this spectrum and using dimensional analysis, it is easy to show that in
the presence of dissipation, the dissipation wavenumber in such a flow is
$k_\eta \sim [\epsilon/(h^2_0 \nu^3)]^{1/5}$, where $\epsilon$ is the mean
energy injection rate.

A scaling compatible with a $\sim k^{-4/3}$ spectrum was observed in laboratory
and field datasets \cite{smith_equilibrium_2003,kaihatu_asymptotic_2007}, where
a spectrum compatible with $\sim k^{-2}$ was also found in shallower regions of
the fluid.

The prediction in Eq.~\eqref{predicc_wt} applies to the BQ model, when
dispersion is not negligible. Before proceeding, we should comment on some
peculiarities of the SW model regarding wave turbulence. First, an inspection of
its dispersion relation, Eq.~\eqref{rel_disp_sw}, indicates that three wave
interactions are possible in this model, and as a result the arguments above for
four-wave interactions do not apply. Weak turbulence theory can be used in
systems with three-waves interactions (with the case of deep water flows being a
paradigmatic one, but see also the case of rotating \cite{galtier_weak_2003} and
of magnetohydronamic \cite{galtier_weak_2000} flows). However, the SW model is
non-dispersive, and as a result the resonance condition is only satisfied for
collinear wave vectors. Resonant interactions can then only couple modes that
propagate in the same direction (i.e., along the ray of the wave), and
non-resonant interactions must be taken into account to consider other
couplings. But more importantly, dispersion is crucial in weak turbulence theory
to have decorrelation between different waves: without dispersion, all modes
propagate with the same velocity, and the modes initially correlated remain
correlated for all times (see, e.g., \cite{lvov_statistical_1997} for a
discussion of these effects in the context of acoustic turbulence).

\begin{figure}
    \centering
    \includegraphics[width=0.48\textwidth]{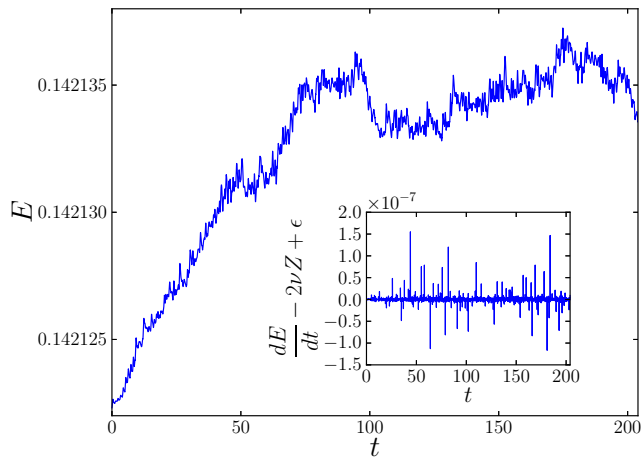}
    \caption{{\it (Color online)} Total energy as a function of time for
        simulation $A06$ (see Table \ref{table1}). As the fluid starts from rest
        and the forcing is applied, energy increases until it reaches a
        turbulent steady state (note energy at $t=0$ is different from zero, as
        the flow potential energy is never zero). All the analysis of the
        simulations was performed after the simulations reached the turbulent
        steady state. {\it Inset:} Energy balance as a function of time (see
        Eq.~\eqref{balance_full}). Note the balance is satisfied up to the
        seventh decimal place.
        }
    \label{eng_cons}
\end{figure}

\section{Numerical simulations} 
\label{simulations}

\begin{table*}
\begin{center}
\begin{ruledtabular}
\begin{tabular}{ c c c c c c c c c }
Simulation & Re & Fr & $D_s$ & $N_l$ & $h_0/L_0$ & $f_0/U_0$ & $[k_{f_1},k_{f_2}]$ & $N$ \\
\colrule
$A01$ & 260 & 0.005 & 0.27 & $1.2\times10^{-5}$ & $8.0\times10^{-4}$ & 0.76 & [3,8] & 1024 \\
$A02$ & 370 & 0.0059 & 0.22 & $1.6\times10^{-5}$ & $6.4\times10^{-4}$ & 0.71 & [3,8] & 1024 \\
$A03$ & 820 & 0.0075 & 0.33 & $2.6\times10^{-5}$ & $4.9\times10^{-4}$ & 0.64 & [3,8] & 2048 \\
$A04$ & 760 & 0.0067 & 0.36 & $2.2\times10^{-5}$ & $5.3\times10^{-4}$ & 0.69 & [3,8] & 2048 \\
$A05$ & 760 & 0.007 & 0.33 & $2.3\times10^{-5}$ & $4.8\times10^{-4}$ & 0.69 & [3,8] & 2048 \\
$A06$ & 360 & 0.0066 & 0.33 & $2.1\times10^{-5}$ & $4.8\times10^{-4}$ & 0.73 & [3,8] & 2048 \\
$A07$ & 570 & 0.0091 & 0.43 & $4.1\times10^{-5}$ & $6.4\times10^{-4}$ & 0.92 & [3,8] & 2048 \\
$A08$ & 350 & 0.0083 & 0.43 & $3.4\times10^{-5}$ & $6.4\times10^{-4}$ & 1 & [3,8] & 2048 \\
$A09$ & 290 & 0.0086 & 0.43 & $3.6\times10^{-5}$ & $6.4\times10^{-4}$ & 0.98 & [3,8] & 2048 \\
$A10$ & 420 & 0.012 & 0.43 & $7.8\times10^{-5}$ & $6.4\times10^{-4}$ & 1.4 & [3,8] & 2048 \\
\end{tabular}
\end{ruledtabular}
\end{center}
\caption{Dimensionless numbers (defined in the text) and parameters for runs in
    set $A$. Re is the Reynolds number, Fr is the Froude number, $D_s$ is the
    dispersivity, $N_l$ is the non-linear number, $h_0/L_0$ is the height of the
    fluid at rest divided by the length of the box, $f_0/U_0$ the amplitude of
    the forcing divided by the rms speed, $k_{f_1}$ and $k_{f_2}$ are
    respectively the minimum and maximum wavenumbers in which the random forcing
    is applied, and $N$ is the linear resolution. In all cases, the Boussinesq
    model was solved.}
\label{table1}
\end{table*}

\begin{table*}
\begin{center}
\begin{ruledtabular}
\begin{tabular}{ c c c c c c c c c }
Simulation & Re & Fr & $D_s$ & $N_l$ & $h_0/L_0$ & $f_0/U_0$ & $[k_{f_1},k_{f_2}]$ & $N$ \\
\colrule
$B01$ & 5600 & 0.022 & 0.14 & $2.5\times10^{-4}$ & $8.0\times10^{-4}$ & 0.45 & [1,5] & 512 \\
$B02$ & 3700 & 0.015 & 0.14 & $1.1\times10^{-4}$ & $8.0\times10^{-4}$ & 0.34 & [1,5] & 512 \\
$B03$ & 5000 & 0.012 & 0.14 & $7.2\times10^{-5}$ & $8.0\times10^{-4}$ & 0.29 & [1,5] & 512 \\
$B04$ & 7100 & 0.013 & 0.11 & $9.1\times10^{-5}$ & $3.2\times10^{-4}$ & 0.24 & [1,5] & 1024 \\
$B05$ & 830 & 0.005 & 0.11 & $1.2\times10^{-5}$ & $3.2\times10^{-4}$ & 0.8 & [3,8] & 1024 \\
$B06$ & 1200 & 0.011 & 0.11 & $6.2\times10^{-5}$ & $3.2\times10^{-4}$ & 0.57 & [3,8] & 1024 \\
$B07$ & 120 & 0.0046 & 0.14 & $1.0\times10^{-5}$ & $8.0\times10^{-4}$ & 0.82 & [3,8] & 512 \\
$B08$ & 980 & 0.012 & 0.17 & $7.9\times10^{-5}$ & $2.5\times10^{-4}$ & 0.54 & [3,8] & 2048 \\
$B09$ & 2500 & 0.038 & 0.27 & $7.4\times10^{-4}$ & $8.0\times10^{-4}$ & 0.31 & [3,8] & 1024 \\
$B10$ & 670 & 0.0042 & 0.17 & $8.5\times10^{-6}$ & $2.5\times10^{-4}$ & 0.79 & [3,8] & 2048 \\
$B_{SW}11$ & 100 & 0.0039 & 0.14 & $7.6\times10^{-6}$ & $8.0\times10^{-4}$ & 0.96 & [3,8] & 512 \\
$B_{SW}12$ & 470 & 0.013 & 0.11 & $9.0\times10^{-5}$ & $3.2\times10^{-4}$ & 0.24 & [1,5] & 1024 \\
\end{tabular}
\end{ruledtabular}
\end{center}
\caption{Dimensionless numbers and parameters for runs in set $B$. Labels are
      as in Table \ref{table1}.  The Boussinesq model was solved in all cases
      except for runs $B_{SW}11$ and $B_{SW}12$, that were done solving the
      shallow water model.}
\label{table2}
\end{table*}

\begin{table*}
\begin{center}
\begin{ruledtabular}
\begin{tabular}{ c c c c c c c c c }
Simulation & Re & Fr & $D_s$ & $N_l$ & $h_0/L_0$ & $f_0/U_0$ & $[k_{f_1},k_{f_2}]$ & $N$ \\
\colrule
$C01$ & 1400 & 0.0067 & 0.27 & $1.9\times10^{-5}$ & $8.0\times10^{-4}$ & 0.56 & [3,8] & 1024 \\
$C02$ & 1400 & 0.0073 & 0.24 & $2.4\times10^{-5}$ & $7.0\times10^{-4}$ & 0.55 & [3,8] & 1024 \\
$C03$ & 1900 & 0.0092 & 0.31 & $4.1\times10^{-5}$ & $4.6\times10^{-4}$ & 0.54 & [3,8] & 2048 \\
$C04$ & 1400 & 0.0057 & 0.43 & $1.6\times10^{-5}$ & $6.4\times10^{-4}$ & 0.74 & [3,8] & 2048 \\
$C05$ & 470 & 0.0067 & 0.54 & $2.2\times10^{-5}$ & $8.0\times10^{-4}$ & 1.1 & [3,8] & 2048 \\
\end{tabular}
\end{ruledtabular}
\end{center}
\caption{Dimensionless numbers and parameters for runs in set $C$. Labels are
      as in Table \ref{table1}. The Boussinesq model was solved in all
      runs.}
\label{table3}
\end{table*}

We performed several numerical simulations of both the shallow water 
and the Boussinesq models. These were done using the GHOST code
\cite{gomez_mhd_2005,gomez_parallel_2005,mininni_hybrid_2011}, which uses a
pseudospectral method with periodic boundary conditions on a 
$L_0 \times L_0 = 2 \pi \times 2 \pi$ sized box (with $L_0$ the box
length), the ``$2/3$ rule'' for the dealiasing
\cite{canuto_spectral_1988}, explicit second order Runge-Kutta for time
stepping, and is parallelized using MPI and OpenMP. Almost all
simulations shown here were done on grids of $N^2=2058^2$ points, with
a few on grids of $N^2=1024^2$ or $512^2$ points (with $N$ the linear 
resolution).  As a result of dealiasing, the maximum resolved wavenumber is 
\begin{equation}
k_\textrm{max}=N/3 .
\label{dealising}
\end{equation}
Note all magnitudes in the code are dimensionless, with the smallest 
wavenumber $k_\textrm{min}=2\pi/L_0=1$, and the largest wavenumber 
$k_\textrm{max}=2\pi/\lambda_\textrm{min}$ being associated with the 
minimum resolved scale $\lambda_\textrm{min}$.

All runs are direct numerical simulations, with all relevant space and time
scales resolved explicitly. The pseudospectral method with the $2/3$ rule is
equivalent to a purely spectral method \cite{canuto_spectral_1988}: it converges
exponentially fast, it conserves all quadratic invariants of the equations
(i.e., there is no numerical dissipation introduced by the method), and it also
has no numerical dispersion. All this was verified explicitly during the
development of the code, using several test problems for the SW and BQ
equations.

Most previous numerical studies on wave turbulence in gravity waves were done at
lower resolutions, with the exception of \cite{yokoyama_statistics_2004}. But
the key difference between previous simulations and the ones presented here
(besides the fact that these are for shallow flows, not for deep flows) is that
the physical model we use does not assume potential flow, and, more importantly,
we do not truncate the non-linear term, thus retaining all high order
non-linearities. Another difference is that we do not introduce an artificial
dissipation term as it is usually done, but one based on physical grounds. The
key motivation for these choices is to be able to compare with experiments in
the future, where vortical motions can develop, and where dissipation also plays
a non-negligible role. To achieve higher resolutions than the ones studied here
becomes increasingly more expensive as the BQ model is dispersive.

All the simulations were started from the fluid at rest. An external
mechanical forcing injected energy in the system, allowing it to reach
for sufficiently long times an out-of-equilibrium turbulent steady
state, after an initial transient. To excite waves, and prevent 
external injection of energy into vortical motions, the forcing had
the following form
\begin{equation}
    \mathbf{F} = \nab f,
\end{equation}
where $f$ is a randomly generated scalar function, with a time correlation
of one time unit, amplitude $f_0$, and applied in a band of wavenumbers in
Fourier space between $k_{f_1}$ and $k_{f_2}$ (see Tables
\ref{table1}, \ref{table2}, and \ref{table3}). Note that having a
mechanical forcing in the momentum equation adds an extra term to
the right hand side of Eq.~\eqref{balance},
\begin{equation}
    \label{balance_full}
    \frac{\mathrm{d} E}{\mathrm{d} t} = - 2 \nu Z + \epsilon,
\end{equation}
where the mean energy injection rate can be computed as
\begin{equation}
    \label{eps_balance}
    \epsilon = \frac{1}{A} \iint\limits_A h \mathbf{u}\cdot \mathbf{f} \mathrm{d}x
    \mathrm{d}y .
\end{equation}

Under the procedure described above, the typical evolution of
  the energy in a numerical simulation is shown in Fig.~\ref{eng_cons}. 
  The energy starts from the value corresponding to the fluid at rest
  (i.e., all the energy is the potential energy associated with the 
  equilibrium height $h_0$). The total energy then grows under the 
  action of the external mechanical forcing, and after $t \approx 80$ the
  system reaches a turbulent steady state in which the energy
  fluctuates around a mean value, and in which the energy injection and
  dissipation are equilibrated on the average. Even though 
  pseudospectral methods are known to introduce no numerical 
  dissipation, in the inset of Fig.~\ref{eng_cons} we also show 
  explicitly that the energy balance (Eq.~\eqref{balance_full}) is
  satisfied with an error of order $10^{-7}$, which remains 
  stable and does not grow even after integrating for very long times.

To ensure that the flow in the simulations remained shallow 
for all excited wavenumbers, we enforced the following condition
\begin{gather}
    \frac{h_0}{\lambda_\textrm{min}} = h_0 \frac{k_\textrm{max}}{2
      \pi} < 1 \nonumber \\
    \Rightarrow h_0 < \frac{6 \pi}{N} .
    \label{cond_disp}
\end{gather}
where $\lambda_\textrm{min}$ is, as already mentioned, the shortest 
wavelength resolved by the code in virtue of the condition given by 
Eq.~\eqref{dealising}.

\section{\label{results} Results}

\subsection{Description and classification of the simulations} 

The spectral behavior of the flow in the simulations depends on the external
parameters.  We can independently control the height of the fluid at rest $h_0$,
the  viscosity $\nu$, the gravity acceleration $g$, the amplitude of the forcing
$f_0$, the range of wavenumbers in which the force is applied, and the linear
resolution $N$. However, all these parameters can be reduced to a smaller set of
dimensionless controlling parameters.

One of these parameters is the Froude number
\begin{equation}
    \textrm{Fr}=\frac{U_0}{\sqrt{g h_0}},
\end{equation}
which measures the ratio of inertia to gravity acceleration in the
momentum equation, and where $U_0$ is the r.m.s.~velocity. 

Another dimensionless parameter is the non-linear number,
$N_l$. In order to be in the regime of weak turbulence, 
nonlinearities should be small. The effect of nonlinearities can 
be measured by how large perturbations in $h$ are compared to 
$h_0$, so we define $N_l$ as
\begin{equation}
    N_l=\frac{h_\textrm{rms}-h_0}{h_0},
\end{equation}
where $h_\textrm{rms}$ is the r.m.s.~value of $h$. 

The two remaining dimensionless numbers are the Reynolds number,
\begin{equation}
    \label{reynolds}
    \textrm{Re} =\frac{U_0 L_f}{\nu},
\end{equation}
where $L_f$ is the forcing scale (defined as $2 \pi/k_{f_0}$), and what 
we will call the dispersivity, $D_s$, defined as
\begin{equation}
    \label{dispersivity}
    D_s = h_0 k_\textrm{max} = \frac{2\pi
        h_0}{\lambda_\textrm{min}} = \frac{N h_0}{6 \pi} 
\end{equation}
following Eq.~\eqref{cond_disp}. This
last number, only relevant for the Boussinesq model, measures how strong the
dispersion is at the smallest scales, and for sufficiently small $D_s$ we 
can expect the solutions of the Boussinesq model to converge to the
solutions of the shallow water model. In fact, it is easy to show from the weak
turbulence spectrum in Eq.~\eqref{predicc_wt} that when the maximum resolved
wavenumber $k_{\textrm{max}}$ is associated with the dissipation wavenumber
$k_\eta$, then
\begin{equation}
    \textrm{Re} \sim \frac{U_0 L_f}{h_0 \epsilon^{1/3}} D^{5/3}_s .
    \label{reycurve}
\end{equation}
Decreasing $D_s$ below the value given by this relation should result in
negligible dispersion at all resolved wavenumbers.  Note that the level of
dispersion in a given Boussinesq run depends on the wavenumber, and $D_s$
actually quantifies the strongest possible dispersion at the smallest scales in
the flow.

By qualitatively assessing each run, we can classify them into three sets, $A$,
$B$, and $C$. In tables \ref{table1}, \ref{table2}, \ref{table3} the different
dimensionless parameters, along with a few other useful quantities, are given
for each simulation in each set, respectively. How and why these three sets
differ from each other will be made clear in the following sections, when we
discuss the actual results. But, for the moment, it is fruitful to analyze the
behavior of the values of $\textrm{Re}$ and $D_s$ in each set, so as to
keep them in mind for later on.

\begin{figure}
    \centering
    \includegraphics[width=0.48\textwidth]{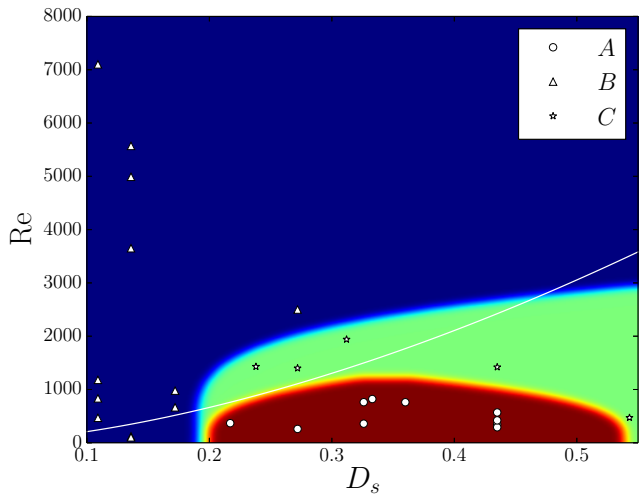}
    \caption{{\it (Color online)} Values of the Reynolds number $\textrm{Re}$
        and dispersivity $D_s$ for all the Boussinesq runs, separated into three
        sets, $A$ (circles in the gray/red region), $B$ (triangles in the
        dark/blue region), and $C$  (stars in the light/green region), according
        to their different spectral behavior as discussed in Sec.~\ref{spectra}.
        The boundaries separating the three regions are arbitrary. The solid
        white curve corresponds to $\textrm{Re} \sim D^{5/3}_s$; points below
        that curve are expected to have non-negligible dispersion (see
        Eq.~\eqref{reycurve}). Note that runs in set $A$ have relatively small
        $\textrm{Re}$ but larger dispersion, while runs in set $B$ have either
        small or negligible dispersion.}
    \label{phase_dia}
\end{figure}

The values of $\textrm{Re}$ and $D_s$ for all the Boussinessq runs are shown in
Fig.~\ref{phase_dia}. As a reference, Fig.~\ref{phase_dia} also shows the curve
given by Eq.~\eqref{reycurve} with $U_0 L_f/(h_0 \epsilon^{1/3})$ estimated from
the values from the simulations in set $A$. Points below that curve are expected
to have non-negligible dispersion. Runs in set $A$ have relatively small
$\textrm{Re}$ ($\lesssim 1000$), and $D_s$ varying between $\approx 0.02$ and
$\approx 0.05$. In other words, dispersion effects in runs in set $A$ are
important. Runs in set $B$ have smaller values of $D_s$ (except for one run with
$D_s \approx 0.27$, all other runs have $D_s <0.2$), and $\textrm{Re}$ varying
between $\approx 100$ and $\approx 7000$.  These runs have small or negligible
dispersion, and note all the SW runs we performed belong to this set. The runs
in set $C$ are intermediate between these two regimes.

Finally, although the mechanical forcing we use introduces no
vorticity in the horizontal velocity field, some vorticity is spontaneously
generated as the flow evolves. This is probably also the case in experiments. In
order to quantify the presence of vortical structures, we calculated the ratio
of vorticity to divergence in the horizontal velocity field
\begin{equation}
    \frac{\langle | \nab \times \mathbf{u} | \rangle }
    {\langle | \nab \cdot \mathbf{u} | \rangle } ,
\end{equation}
which turns out to be $\approx 0.1$ for all simulations. As a result,
although the flow is not perfectly irrotational, the amplitude of
vortical modes is small compared with the amplitude of modes
associated with the waves.

\subsection{Energy spectra} 
\label{spectra}

\begin{figure}
    \centering
    \includegraphics[width=0.48\textwidth]{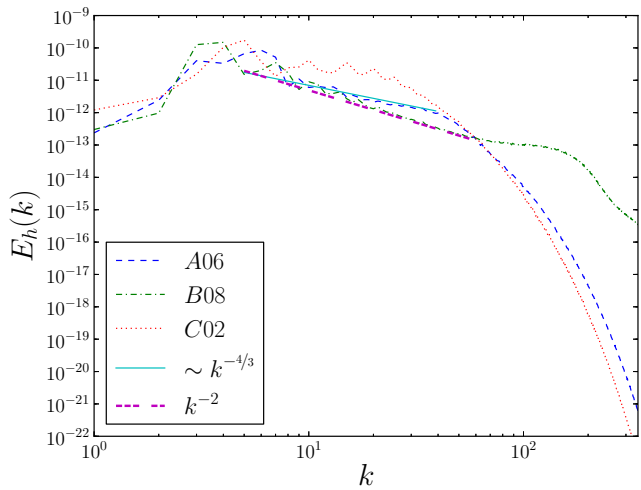}
    \caption{{\it (Color online)} Power spectrum of $h$ (proportional to
        the spectrum of the potential energy) for runs $A06$ (BQ model, $2048^2$
        grid points, $\textrm{Re} = 360$, and $D_s = 0.33$),
        $B08$ (BQ model, $2048^2$ grid points, $\textrm{Re} = 980$,
        and $D_s = 0.17$), and $C02$ (BQ model, $1024^2$ grid
        points, $\textrm{Re} = 1430$, and $D_s = 0.24$). Two
        power laws, $\sim k^{-4/3}$ and $\sim k^{-2}$, are shown as references.}
    \label{spectra_far}
\end{figure}

\begin{figure}
    \centering
    \includegraphics[width=0.48\textwidth]{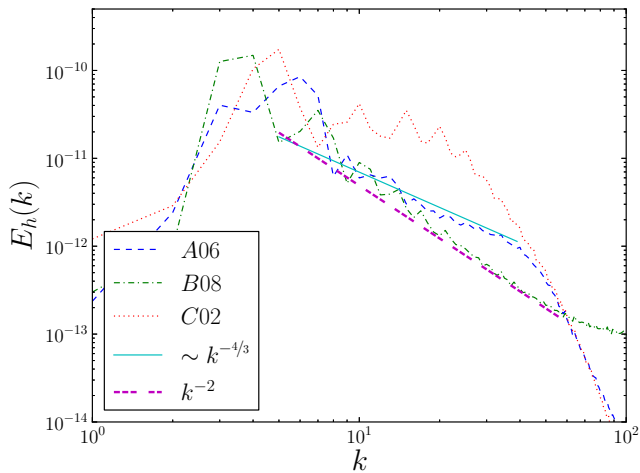}
    \caption{{\it (Color online)} Detail of the three spectra
      in Fig.~\ref{spectra_far} for a subset of wavenumbers to show 
      the inertial range of the runs. Note the scaling of runs $A06$ 
      and $B08$.}
    \label{spectra_zoom}
\end{figure}

The power spectrum of $h$ (proportional to the spectrum of the potential energy)
as a function of the wave number is shown in Fig.~\ref{spectra_far} for runs
$A06$, $B08$, and $C02$. Figure \ref{spectra_zoom} shows a close-up of the same
spectrum in the inertial range. It is clearly seen that runs in each set show a
different behavior. On the one hand, the run belonging to group $A$ has an
inertial range compatible with $\sim k^{-4/3}$ scaling, which is the spectra
predicted by weak turbulence. On the other hand, the run in set $B$ displays an
inertial range compatible with $\sim k^{-2}$ dependency. While this spectrum is
not predicted by weak turbulence, it was observed before in experiments and
observations \cite{kaihatu_asymptotic_2007}.  The run in group $C$ shows a
shallower spectrum with no clear inertial range. We think of runs in this set as
transitional between the other two.

\begin{figure}
    \centering
    \includegraphics[width=0.48\textwidth]{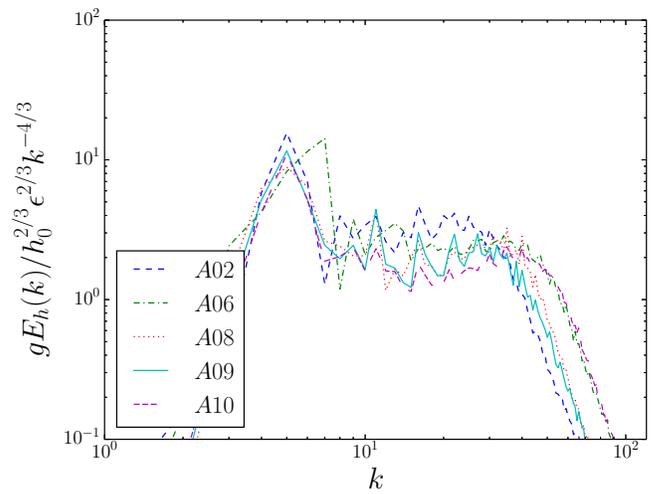}
    \caption{{\it (Color online)} Compensated spectrum of potential energy
        for several simulations in set $A$. The spectra are compensated by
        $h^{2/3}_0 \epsilon^{2/3} k^{-4/3}$. The average slope for all
        the runs is $-1.34\pm0.12$.}
    \label{compensado_A}
\end{figure}

\begin{figure}
    \centering
    \includegraphics[width=0.48\textwidth]{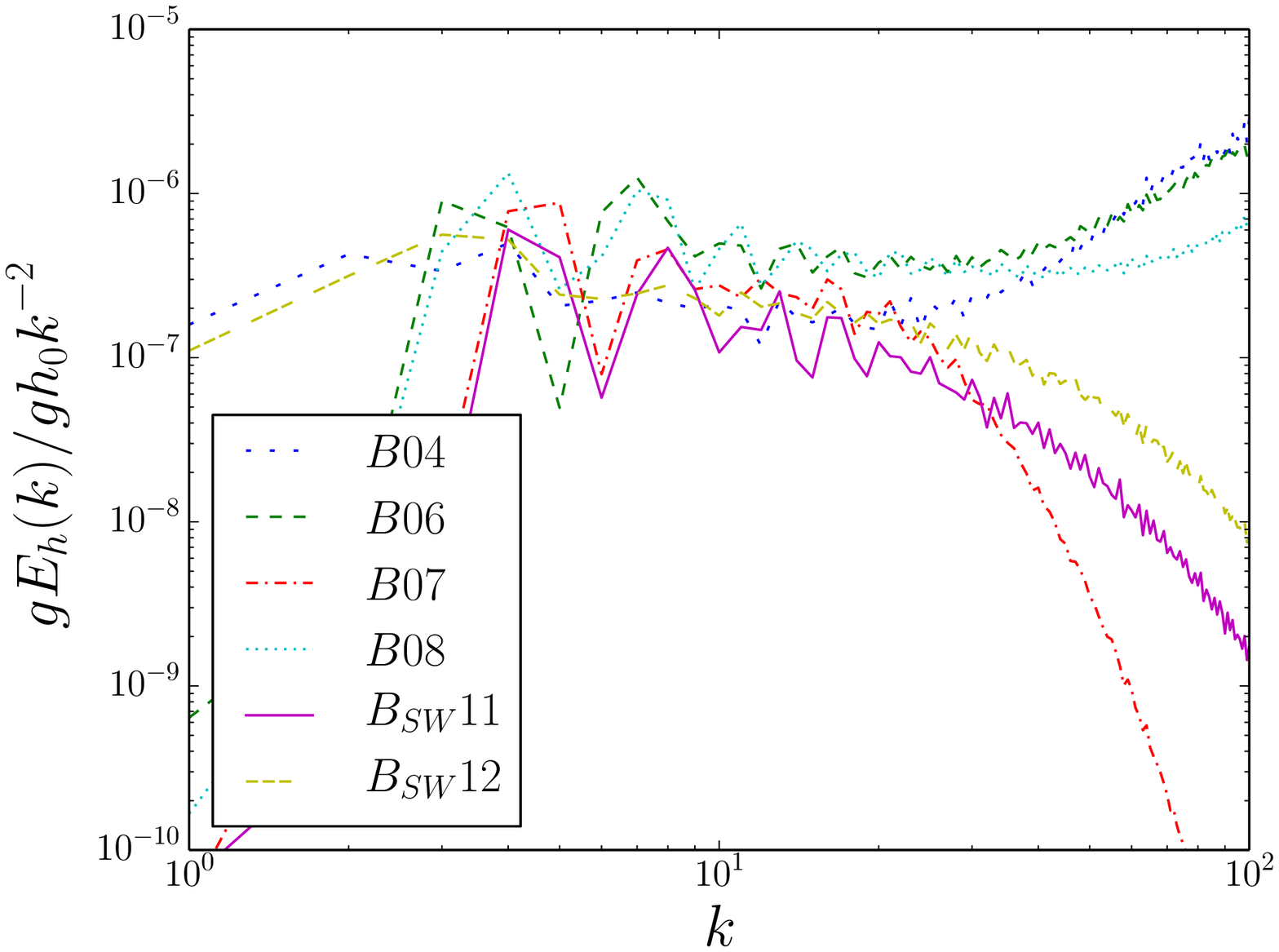}
    \caption{{\it (Color online)} Compensated spectrum of potential energy
        for several simulations in set $B$. The spectra are compensated by
        $g h_0 k^{-2}$. The average slope for all the runs is $-2.18\pm0.29$.}
    \label{compensado_B}
\end{figure}

\begin{figure}
    \centering
    \includegraphics[width=0.48\textwidth]{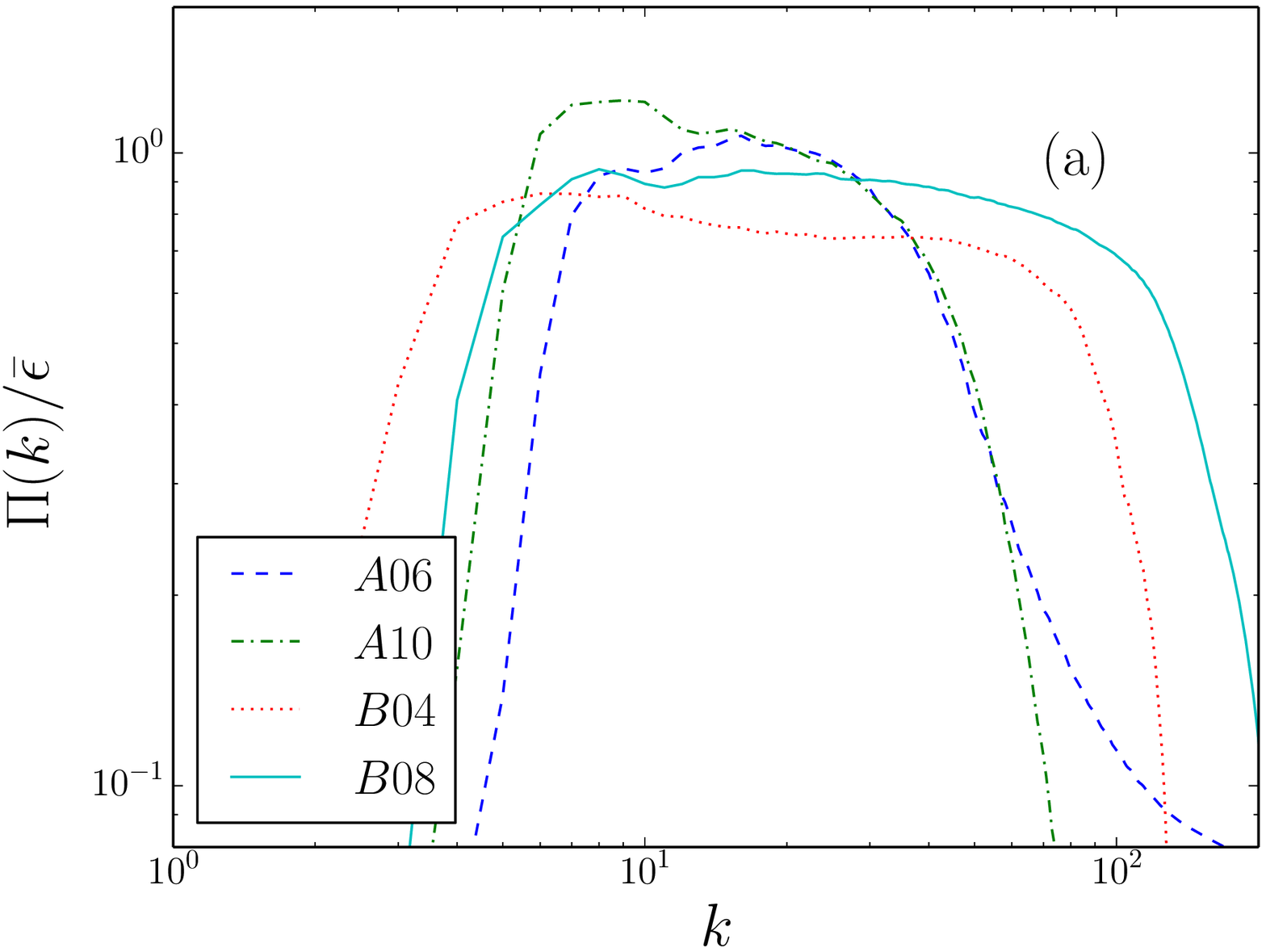} \\
    \includegraphics[width=0.48\textwidth]{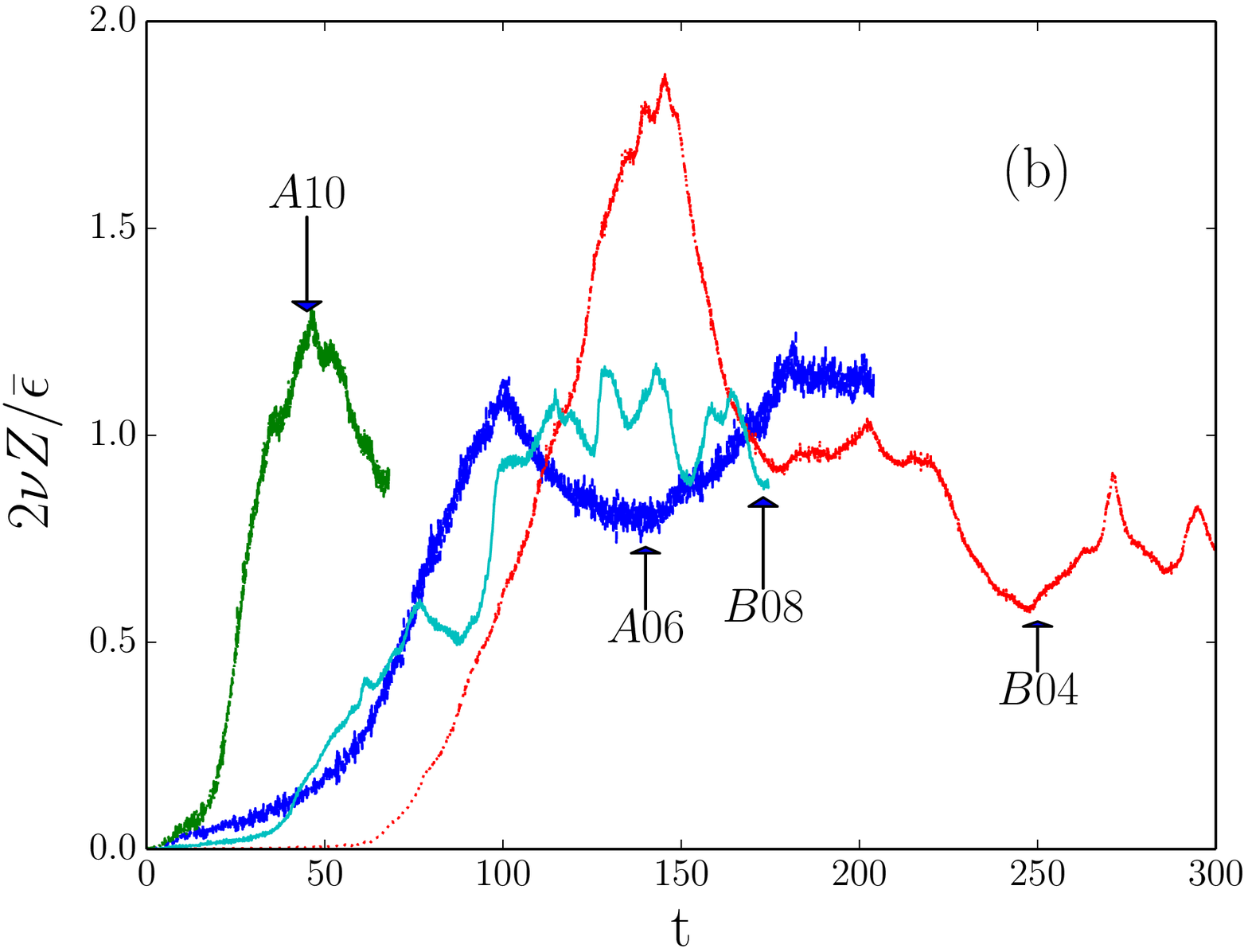}
    \caption{{\it (Color online)} (a) Energy flux (normalized by the mean
        energy injection rate) as a function of $k$. For each simulation, a
        range of wavenumbers can be identified for which $\Pi(k)$ remains
        approximately constant, and this range is in reasonably good agreement
        with the inertial ranges identified in Figs.~\ref{compensado_A} and
        \ref{compensado_B}. (b) Energy dissipation rate (normalized by the
        averaged in time energy injection rate) as a function of time for the
        same simulations.
      } 
    \label{flux}
\end{figure}

The other runs in sets $A$, $B$, and $C$ show similar power spectra for $h$.  To
show this, we present the compensated spectra for the simulations in sets $A$
and $B$ in Figs.~\ref{compensado_A} and \ref{compensado_B} respectively
(simulations from set $C$ do not have a clearly defined inertial range and are
therefore not shown). The simulations from set $A$ are compensated by $h^{2/3}_0
\epsilon^{2/3} k^{-4/3}$ (which is the weak turbulence spectrum, using the
height of the fluid column at rest, $h_0$, and the energy injection rate,
$\epsilon$, as prefactors), while the ones in set $B$ are compensated by $g h_0
k^{-2}$ (more details on the choice of the prefactor are given below). These
figures indicate that, within statistical uncertainties, all spectra in each set
collapse to the same power laws, and that the simulations are well converged
from the point of view of spatial resolution. Furthermore, we verified that the
energy flux is approximately constant in the scales corresponding to the
inertial range of each simulation. Within the limitations of spatial resolution
and the drop in the flux for large wavenumbers caused  by viscous dissipation,
an incipient inertial range can be identified in the flux of each simulation.
Figure~\ref{flux} shows the instantaneous energy flux (normalized by the energy
injection rate $\overline{\epsilon}$ averaged over time) as a function of $k$
for several simulations in sets $A$ and $B$. The energy flux $\Pi(k)$ was
calculated from the energy balance equation in Fourier space, as is usually done
for turbulent flows. Figure~\ref{flux} also shows the normalized energy
dissipation rate as a function of time (equivalent to the normalized energy flux
as a function of time) for the same runs, to show that this quantity fluctuates
around a mean value in the turbulent steady state.

The kinetic energy spectrum is similar to the power spectrum of $h$, and in
approximate equipartition with the potential energy spectrum once the system
reaches a turbulent steady state.  It is interesting to analyse this in the
light of the values of the dimensionless parameters in the runs as shown in
Fig.~\ref{phase_dia}. As was explained in the previous section, set $A$
    corresponds to runs with lower Reynolds number and larger dispersivity
    ($\textrm{Re}\lesssim 1000$, and $D_s$ varying between $\approx
    0.02$ and $\approx 0.05$). As a result, these runs can be expected to
    display weak turbulence behavior as described in Section~\ref{sec:weak},
    because the nonlinearities are not so large as to break the weak
    turbulence hypothesis \cite{onorato_four-wave_2008}, and the dispersion is
    not so low as to render the higher order terms of Eq.~\eqref{rel_disp_bous}
    negligible (in which case four-wave interactions would no longer be 
    dominant, and the hypothesis used to derive
    Eq.~\eqref{kinetic_eq} would not be satisfied). In
    contrast, runs in set $B$ have larger $\textrm{Re}$ and lower
    $D_s$ (except for one run with $D_s \approx
    0.27$, all other runs have $D_s <0.2$, and $\textrm{Re}$ varying
    between $\approx 100$ and $\approx 7000$). In this case dispersion is
    smaller or negligible, while nonlinearities can be expected to be larger, two
    conditions that render the derivation resulting in Eq.~\eqref{predicc_wt}
    invalid.

\begin{figure}
    \centering
    \includegraphics[width=0.48\textwidth]{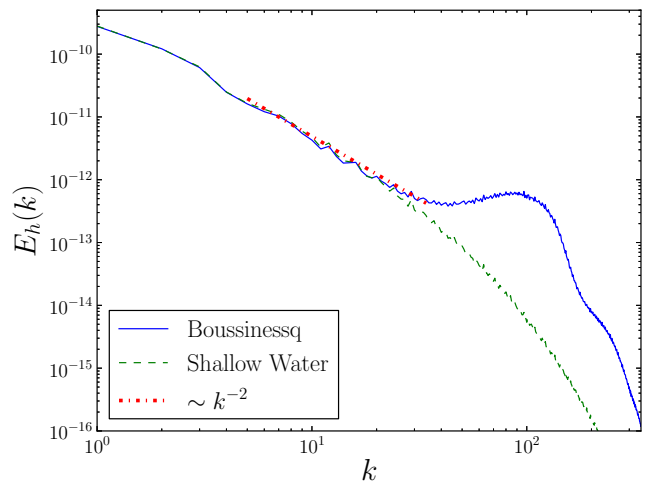}
    \caption{{\it (Color online)} Power spectrum of $h$
      for simulations $B04$ and $B_{SW}12$. The former corresponds 
      to a numerical solution of the BQ model, while the later to a 
      solution of the SW model. A $\sim k^{-2}$ power law is 
      indicated as a reference.}
    \label{comp_sp}
\end{figure}

Moreover, the $\sim g h_0 k^{-2}$ spectra observed in the simulations in set $B$
include those runs that solve the SW model. Therefore, these spectra cannot be
explained by weak turbulence, as SW simulations have no dispersion and the
arguments in Section \ref{sec:weak} do not apply. Also, note that in the
non-dispersive limit, for constant and fixed $h$, the SW equations can be
reduced to the two-dimensional Burgers equations, which amplify negative field
gradients by strong nonlinearities resulting in sharp fronts in the velocity.
Such a field would actually have a spectrum $\sim k^{-2}$ (note
  this is also the behavior expected for two-dimensional
  non-dispersive acoustic turbulence \cite{lvov_statistical_1997},
  that also develops sharp fronts). The spectrum can be obtained from 
dimensional analysis and the scaling that results for the energy is
equivalent to Phillips' spectrum \cite{phillips_equilibrium_1958} but in two
dimensions. In the presence of strong nonlinearities, we can assume 
that the nonlinear and gravity terms are of the same order,
\begin{equation}
    \mathbf{u} \cdot \nab \mathbf{u} \sim g \nab h . 
\end{equation}
It is also reasonable to assume that the kinetic and potential energies will be
of the same order (i.e., in equipartition) in the turbulent steady state. This
implies that $g$ is the only dimensional constant the spectra can depend on.
This is precisely how Phillips derived his spectrum. 

With these assumptions in mind, it is easy to obtain the observed spectra. The
energy spectrum has units of energy in the fluid column per unit surface per
wavenumber, $E(k) \sim h_0 u^2/k$, and assuming $E(k) \sim g h_0 k^{-\alpha}$,
from dimensional analysis the only possible solution is
\begin{equation}
    E(k) \sim g h_0 k^{-2} .
\end{equation}
The independence of the spectrum on the energy injection rate suggests that the
energy transfer between the different scales must take place by a mechanism such
as wave breaking in the case of Phillips' spectrum, which occurs when the slope
of the surface is larger than a critical value, or by nonlinear wave steepening
in our case (which is finally regularized by the viscosity). Such a mechanism is
independent of the power injected by external forces. Of course, this can only
hold in a region of parameter space, as in the presence of weak forcing and
dispersion, the solution in Eq.~\eqref{predicc_wt} is expected instead.

In summary, based on the numerical results, the simulations with weaker forcing
and higher dispersion develop a spectrum compatible with the predictions from
weak turbulence theory, while the runs with stronger forcing or with less (or
no) dispersion are compatible with dimensional analysis based on strong
turbulence arguments.

\subsection{Comparison between SW and BQ models} 
\label{comp_sw_bq}

\begin{figure}
    \centering
    \includegraphics[width=0.48\textwidth]{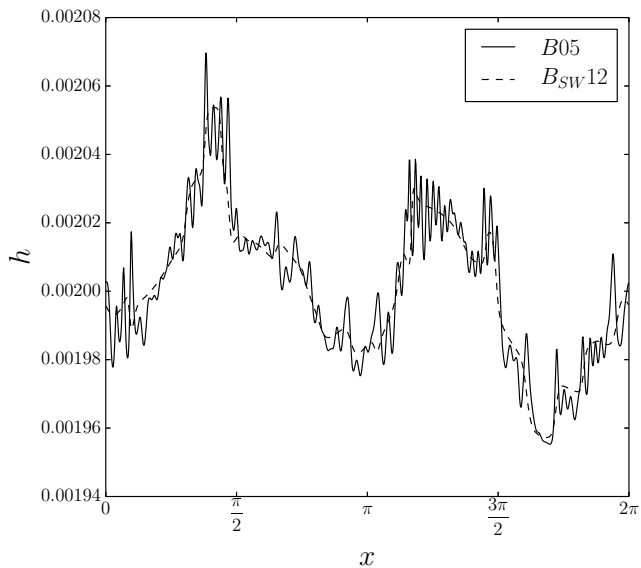}
    \caption{One dimensional cut of the height $h$ in the turbulent
      steady state of runs $B04$ and $B_{SW}12$, at the same
      time. The former run corresponds to a numerical solution of the 
      BQ model, while the latter to a solution of the SW model. While
      the long length scales show the same behavior in both runs, note
      the BQ model has larger fluctuations at short length scales. Both
      runs were computed with a linear resolution of $N=1024$ grid
      points, and the fast fluctuations are well resolved.}
    \label{secc20}
\end{figure}

All simulations of the SW model belong to set $B$, as that is the set of runs
that has negligible or no dispersion. All other sets have moderate dispersion,
and as a result the flow dynamics cannot be captured by the SW model. Note that
runs in set $B$ are also the runs with an inertial range compatible with $\sim
k^{-2}$ scaling. However, the BQ and SW runs in set $B$ are not identical. In
this subsection we discuss the differences between these runs.

As an example of two runs with and without dispersive effects, the power spectra
of $h$ for runs $B04$ and $B_{SW}12$ are shown in Fig.~\ref{comp_sp}.  Both
simulations have the same parameters, except for the viscosity which is larger
in the simulation using the SW model.  At small wavenumbers, where dispersion is
negligible, the spectra of the BQ and SW models coincides. For  wavenumbers
larger than $\approx 30$, dispersion in the BQ model becomes important and a
bump (an accumulation of energy at small scales) develops. This accumulation in
the BQ model results in an increased dissipation (as dissipation is proportional
to $k^2 E(k)$), thus allowing us to simulate the system with smaller viscosity.
This difference at large wavenumbers is the most distinct feature in the two
spectra in Fig.~\ref{comp_sp}.

As a result of the extra power at larger wavenumbers, dispersion in the BQ model
results in more prominent small scale features, and in rapidly varying waves. As
an example, Fig.~\ref{secc20} shows a transversal cut in the elevation field for
runs $B04$ and $B_{SW}12$. The cuts are taken at the same place and at the same
time in both runs. Even though both simulations have the same behavior at large
scales, at short length scales the BQ model presents fast fluctuations. These
fluctuations are well resolved (the cut corresponds to $1024$ grid points), and
there is no indication that resolution is insufficient to resolve the sharp
gradients. In the BQ model, while the large scales may correspond to a shallow
flow, as long as there is enough scale separation, there will always be a
wavenumber where the finite depth effects can be seen. Thus, the Boussinesq
equations provide an interesting model to study weakly dispersive waves.

Regarding the accumulation of energy that leads to a flatter spectrum for high
wavenumbers in some of the BQ simulations (for several runs in set $B$ as can be
seen in Fig.~\ref{compensado_B}, but specially in the runs in set $C$), such an
accumulation has been observed before in turbulent flows. As mentioned above, we
verified that this accumulation is not the result of insufficient resolution
(e.g., by comparing the runs with different grid points $N$). The accumulation
of energy in the spectrum near the dissipative range is often termed
``bottleneck'', and bottlenecks can have dissipative
\cite{falkovich_bottleneck_1994} or dispersive
\cite{graham_highly_2007,krstulovic_dispersive_2011} origins.  In the former
case, the accumulation results from the viscous damping of the triads at small
scales, resulting in a decrease of the energy flux.  Such a viscous bottleneck
should be visible also in the non-dispersive simulations, and its absence in
those runs indicates a dispersive origin. In the latter case, the bottleneck
arises from the increasingly harder to satisfy resonant condition for the wave
frequencies, as the waves become faster at smaller scales. Models with a field
filtered by the Helmholtz operator (as is the case for the BQ model, see
Eq.~\ref{eq:Helmholtz}) tend to develop a bottleneck (see
\cite{graham_highly_2007} for a detailed description of its origin). A
qualitative way to explain the tendency towards a flatter spectrum in the BQ
model can be obtained by assuming that dispersion is strong enough for the
dispersive term to be balanced with the buoyancy and with the non-linear terms
in the BQ equations (i.e., all terms are of the same order). Then the energy
spectra can depend only on both $g$ and $h_0$, and a possible solution is $E(k)
\sim g h_0^2$.  A detailed study of the origin of this bottleneck is left for
future work.

At this point it is worth pointing out that when $D_s \approx 1$ and
dispersion becomes too strong, the Boussinesq approximation breaks down as more
terms in the Taylor expansion in Eq.~\eqref{taylor_exp} should be preserved. As
a result, the Boussinesq approximation is useful as long as $D_s < 1$ at
the smallest excited scales in the system.  On the other hand, from
Fig.~\ref{phase_dia}, if $D_s \lesssim 0.15$ the behavior of the system
in the inertial range is that of a shallow water flow for all Reynolds numbers
studied.

\begin{figure}
    \includegraphics[width=0.48\textwidth]{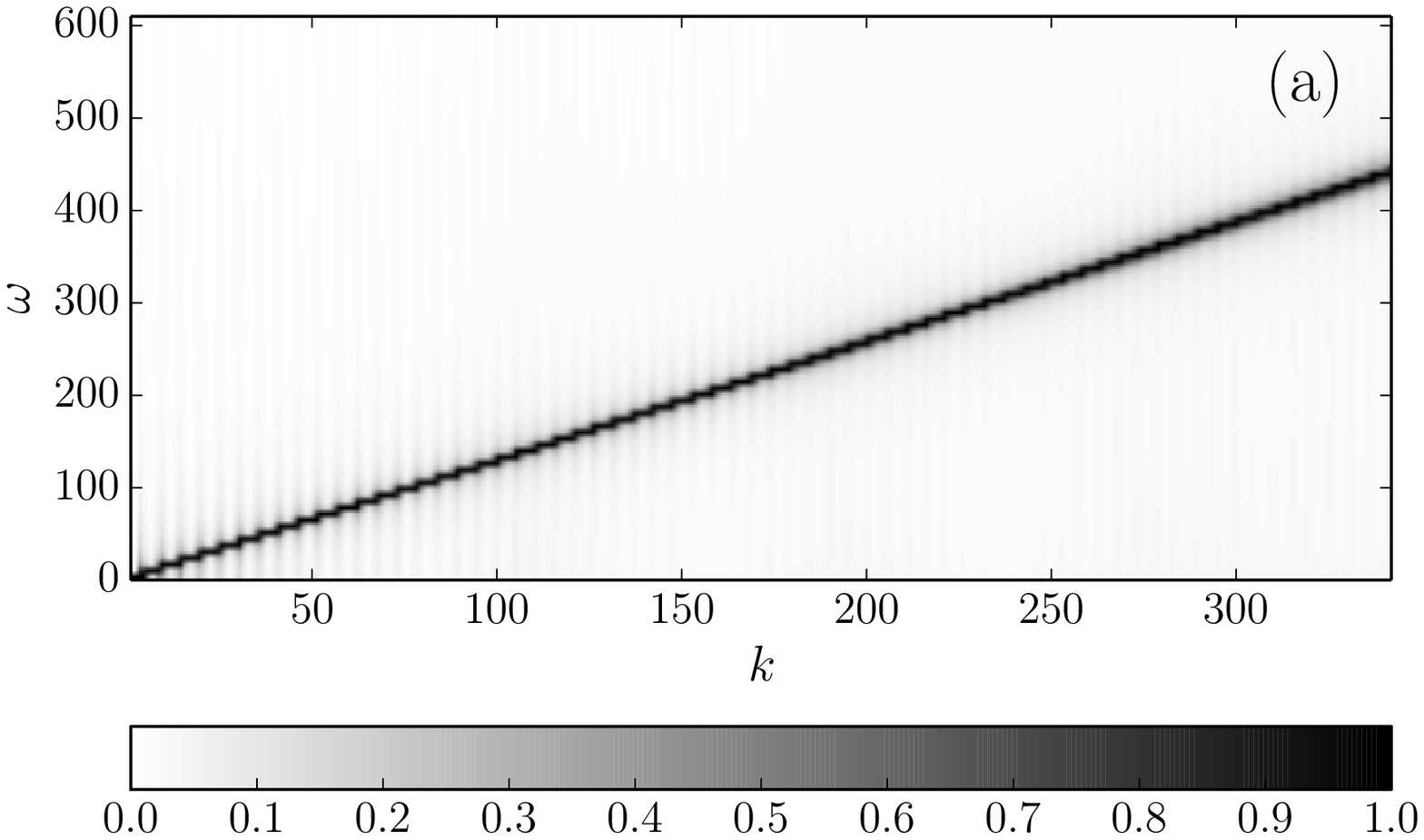}
    \includegraphics[width=0.48\textwidth]{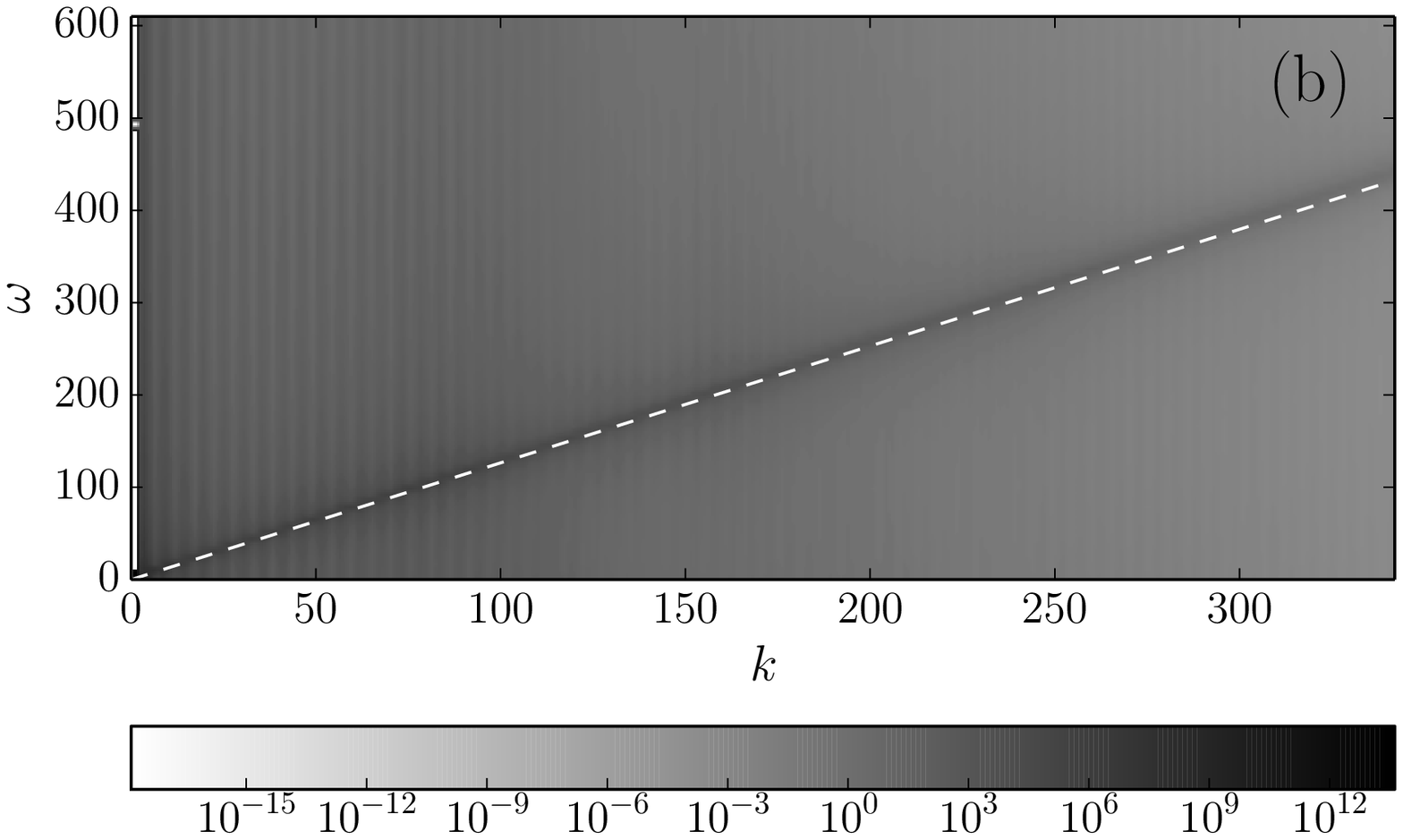}
    \caption{Power spectrum $E_h(k,\omega)$ for simulation $B04$. The darker
        regions correspond to larger power density, while the lighter regions
        correspond to smaller power density. (a) Normalized power
        spectrum $E_h(k,\omega)/E_h(k)$. (b) Non-normalized power
        spectrum. The white dashed line appearing in the bottom panel
        indicates the linear dispersion relation from Eq.~\eqref{rel_disp_bous}.
        Note that as in this run dispersion is negligible, the dispersion
        relation is almost that given by Eq.~\eqref{rel_disp_sw}, and
        non-dispersive.} \label{cubo32}
\end{figure}

\subsection{Time-resolved spectra and non-linear dispersion relations} 

Wavenumber spectra, as the spectra discussed so far, give information of how
energy is distributed in spatial scales, but do not provide a quantitative
estimate of how much energy in the system is associated with wave motions. A
frequency spectrum $E(\omega)$ is often obtained from the wavenumber spectrum $E(k)$
using the dispersion relation (or vice versa). However, in systems that can sustain both wave
and vortical motions there is no clear justification to use the dispersion
relation to go from one spectrum to the other.

A quantification of the amount of energy in waves, and on whether non-linear
effects change the dispersion relation of the system from the linear one, can be
directly obtained from the frequency and wavenumber spectrum $E(k, \omega)$
without any assumption. The spectrum $E(k, \omega)$ can be computed by storing
the Fourier coefficients of the height $\hat{h}({\bf k}, t)$ as a function of
time (as well as the Fourier coefficients of the velocity field), then computing
the Fourier transform in time, and finally computing the isotropic power
spectrum by averaging in the $(k_x, k_y)$-plane.  To this end, several
large-scale wave periods and turnover times must be stored (to resolve the
slowest frequencies in the system), with sufficient time resolution $\Delta t$
to resolve the fastest frequencies.  In the analysis we show below, time series
spanning at least three periods of the slowest waves were used, and with time
resolution $\Delta t \approx 3 \times 10^{-4}$.

\begin{figure}
    \includegraphics[width=0.48\textwidth]{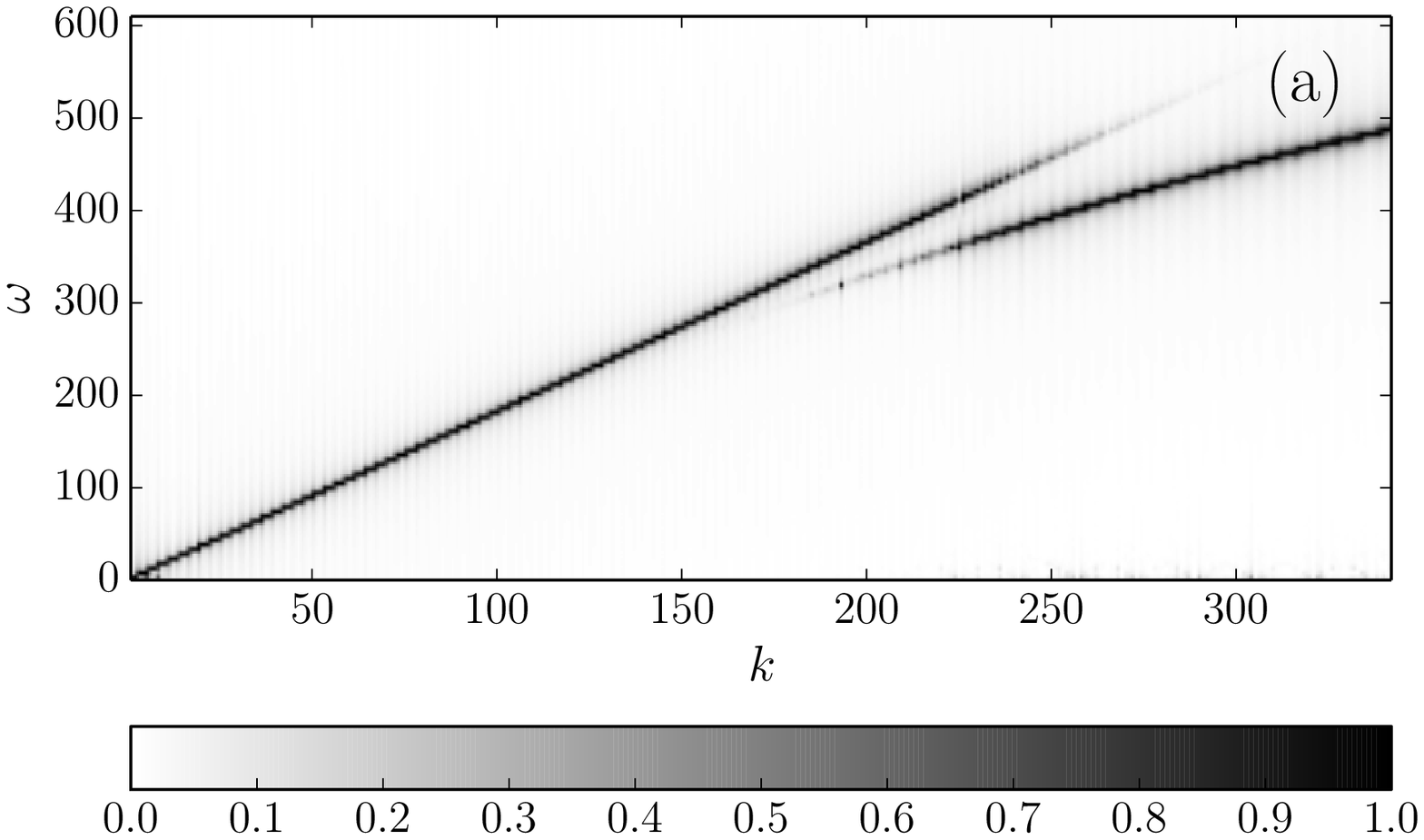}
    \includegraphics[width=0.48\textwidth]{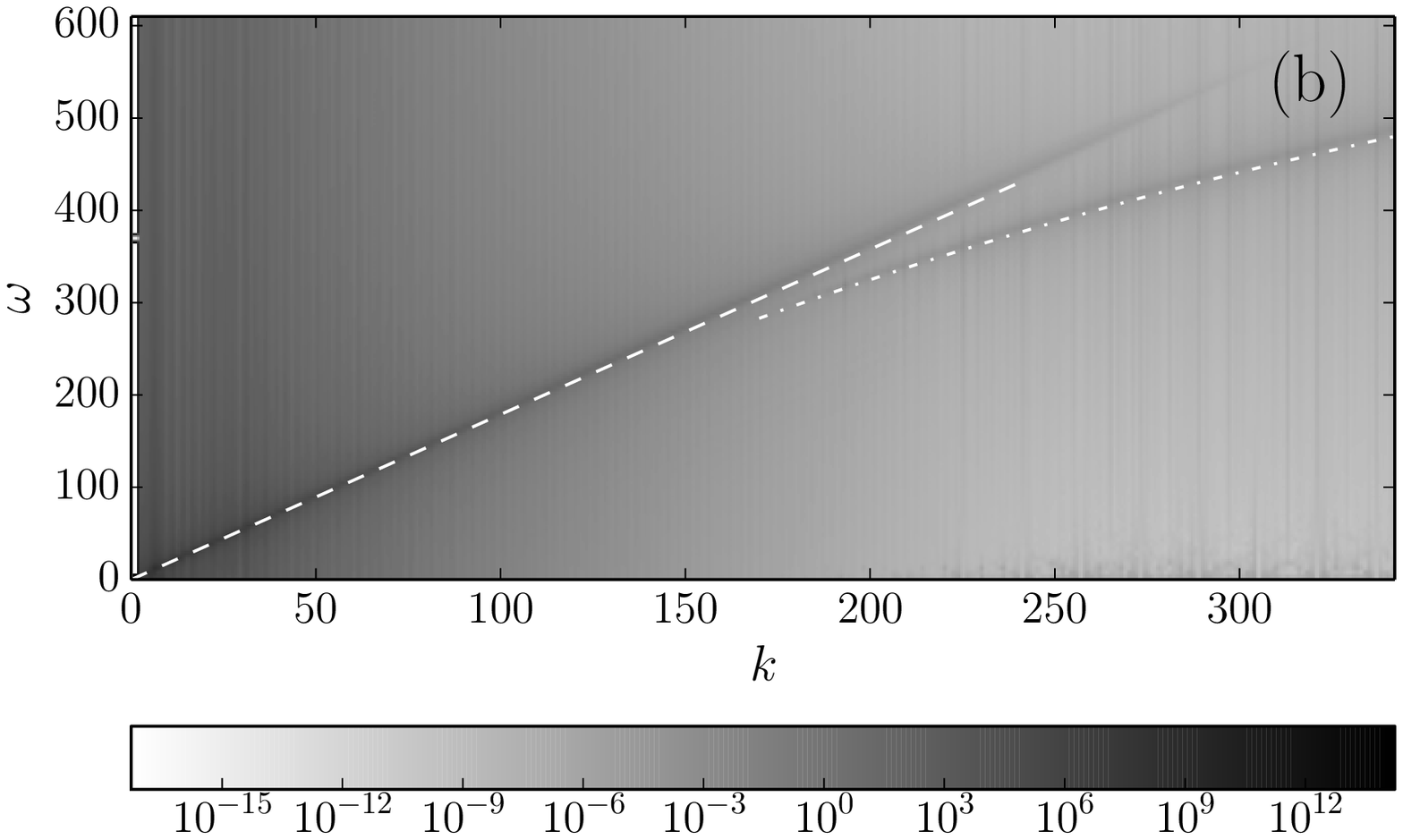}
    \caption{Power spectrum $E_h(k,\omega)$ for simulation $A02$. The darker
        regions correspond to larger power density, while the lighter regions
        correspond to smaller power density. (a) Normalized power
        spectrum. (b) Non-normalized power spectrum. The white dashed
        line appearing in the bottom panel indicates the (non-dispersive)
        linear dispersion relation from Eq.~\eqref{rel_disp_sw}, and the white
        dash-dotted line indicates the BQ dispersion relation from
        Eq.~\eqref{rel_disp_bous}.}
    \label{cubo20}
\end{figure}

\begin{figure}
    \centering
    \includegraphics[width=0.48\textwidth]{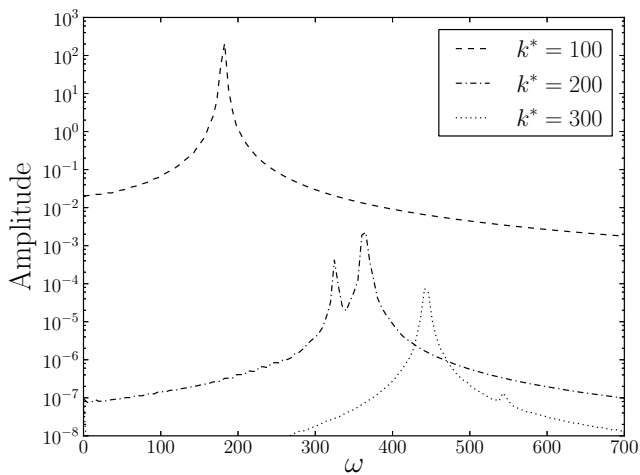}
    \caption{Cross sections of $E_h(k,\omega)$ at different (and
      fixed) values of $k=k^*$ for run $A02$. Note the peaks and 
      surrounding wavenumbers have most of the power. Note also 
      the two peaks for $k=200$, one corresponding to the 
      shallow-water dispersion relation, and the other to the 
      Boussinesq dispersion relation. }
   \label{blnc}
\end{figure}

Figures \ref{cubo32} and \ref{cubo20} show the power spectrum of the flow height
$E_h(k,\omega)$ for simulations $B04$ and $A02$ respectively. The linear
dispersion relations for shallow water flows (Eq.~\ref{rel_disp_sw}) and for
Boussinesq flows (Eq.~\ref{rel_disp_bous}) are also shown as references, using
the parameters from each run. Note both runs present an energy accumulation near
the dispersion relation. This indicates most of the energy is in the waves, and
remains there as time evolves. As we are not solving the equations for a
potential flow, and the system can develop vortical motions, this tells us that
the non-linear energy transfer is mostly done between waves, and that the energy
injected at large scales in wave motions is mostly transferred towards wave
motions at smaller scales and faster frequencies. This is needed for weak
turbulence to hold, but is also observed in run $B04$ that has a spectrum
compatible with strong turbulence phenomenological arguments.  There is also a
turbulent broadening of the dispersion relation, also visible in cross sections
of the spectrum at different wavenumbers in Fig.~\ref{blnc}. From this
broadening, the characteristic time of non-linear wave interactions can be
obtained, as was done in \cite{miquel_nonlinear_2011}.

Some of the most important results in this paper are associated with these two
figures. First, note that in run $B04$ most of the energy is concentrated near a
dispersion relation that, as dispersion is negligible, corresponds in practice
to the non-dispersive shallow-water case (Eq.~\ref{rel_disp_sw}). All runs in
set $B$ have the same spectral behavior in $E_h(k,\omega)$, and confirm that the
$\sim k^{-2}$ spectrum is observed when dispersion is negligible or absent
(i.e., when the flow is sufficiently shallow). Second, note that the spectrum
$E_h(k,\omega)$ in run $A02$ presents clear signs of dispersive effects (i.e.,
most of the energy for large enough $k$ is concentrated over a curve that
deviates from a linear relation between $k$ and $\omega$), and this run displays
a scaling in $E_h(k)$ compatible with the weak turbulence prediction $\sim
k^{-4/3}$. This behavior was observed in the other runs in set $A$.

However, $E_h(k,\omega)$ for runs in set $A$ presents yet another interesting
feature. As expected, for small $k$ the dispersion is negligible and the energy
is concentrated over a straight line in $(k,\omega)$ space. At large $k$, as
already mentioned, the effective dispersion relation is compatible with that of
the linearized Boussinesq equations. But at intermediate wavenumbers two
branches of the dispersion relation can be observed, one that is compatible with
non-dispersive waves and another compatible with dispersive waves.  When both
branches are present, their amplitudes are of the same order, as can be seen in
Fig.~\ref{blnc}.

At first sight, the existence of these two branches could be 
  attributed to bound waves. Bound waves are small amplitude waves 
  which are {\it bounded} to a parent wave of larger amplitude. The 
  waves are bounded in the sense that they follow the parent wave,
  i.e., they travel with the same phase velocity as the parent, and 
  thus they follow an anomalous dispersion relation (see, e.g., a 
  discussion of bound waves in the context of gravito-capillary waves in 
  \cite{longuet-higgins_generation_1963,herbert_observation_2010}). The 
  condition that they have the same phase velocity as the parent wave
  implies that they must follow a modified dispersion relation which
  verifies $\Omega (k) = \omega (k_0) k/k_0$, where $k_0$ is the
  wavenumber of the parent wave. Bound waves result in multiple
  branches in the $E(k,\omega)$ spectrum (and in multiple peaks in the 
  frequency spectrum). Indeed, it is easy to show that for $k=Nk_0$, 
  $N=2,3,4,\dots$, these multiple branches satisfy
\begin{equation}
\Omega_N (Nk_0) = N\omega (k_0) 
\label{bound_condition}
\end{equation}
(see, e.g., the discussion in \cite{herbert_observation_2010}). 
Extending the analysis in \cite{herbert_observation_2010} to 
our case, bound waves in the BQ model should satisfy the 
following dispersion relation,
\begin{equation}
    \label{boundwaves}
    \Omega_N (Nk) = \frac{c_0 k}{\sqrt{1 + \frac{h^2_0 k^2}{3 N^2}}} ,
\end{equation}
which verifies Eq.~\eqref{bound_condition}. However, the
second branch in Fig.~\ref{cubo20} cannot be described by this 
dispersion relation for any value of $N$ up to 4, and thus they are
not bound waves in the sense often used in oceanography.

Another explanation for the existence of these two branches can be given by keeping
in mind that at intermediate wavenumbers slight variations in the fluid depth
may trigger a transition in the waves from dispersive to non-dispersive (as the
level of dispersion depends on the product of the wavenumber with the surface
height).  Indeed, in the turbulent flow there are waves with short wavelengths
which ride over long ones, that have a larger amplitude. For sufficient scale
separation, the fast waves see an effective depth that can be larger or smaller
than $h_0$ depending on whether the wave is on a crest or a valley of the slow
wave, generating in one case dispersive waves, and in the other non-dispersive
waves.

We can estimate the variation in the effective dispersion at a given wavenumber
$k$. In simulation $A02$, $h_0=4\times 10^{-3}$ and the longer waves have an
amplitude $\delta \approx 4\times 10^{-5}$ (as can be estimated, e.g., from the
maximum value of the power spectrum of $h$). From the system dispersion
relation,
\begin{equation}
    \omega^2 = c^2_0 k^2 \left(1 - \frac{1}{3} h_0^2 k^2 \right) ,
\end{equation}
dispersion is controlled by the amplitude of the $h_0^2 k^2/3$ term.  Assuming
that fast waves experience an effective depth $h_0 \pm \delta$ (where the sign
depends on whether they are on a valley or a crest), the variation in the
dispersion is proportional to the difference between $(h_0-\delta)^2$ and
$(h_0+\delta)^2$. So, for this simulation, the variation is around $4\%$, and
when multiplied by $k^2$, it is sufficient to explain the two branches in
$E_h(k,\omega)$ for $k$ between $\approx 150$ and $250$.

\subsection{Time frequency energy spectra} 

From the spectra in Figs.~\ref{cubo32} and \ref{cubo20} the
  frequency spectrum $E_h(\omega)$ can be easily
  obtained, simply by summing over all wavenumbers,
\begin{equation}
E_h(\omega) = \sum_k E_h(k,\omega) .
\label{freqspectra}
\end{equation}

\begin{figure}[H]
    \centering
    \includegraphics[width=0.48\textwidth]{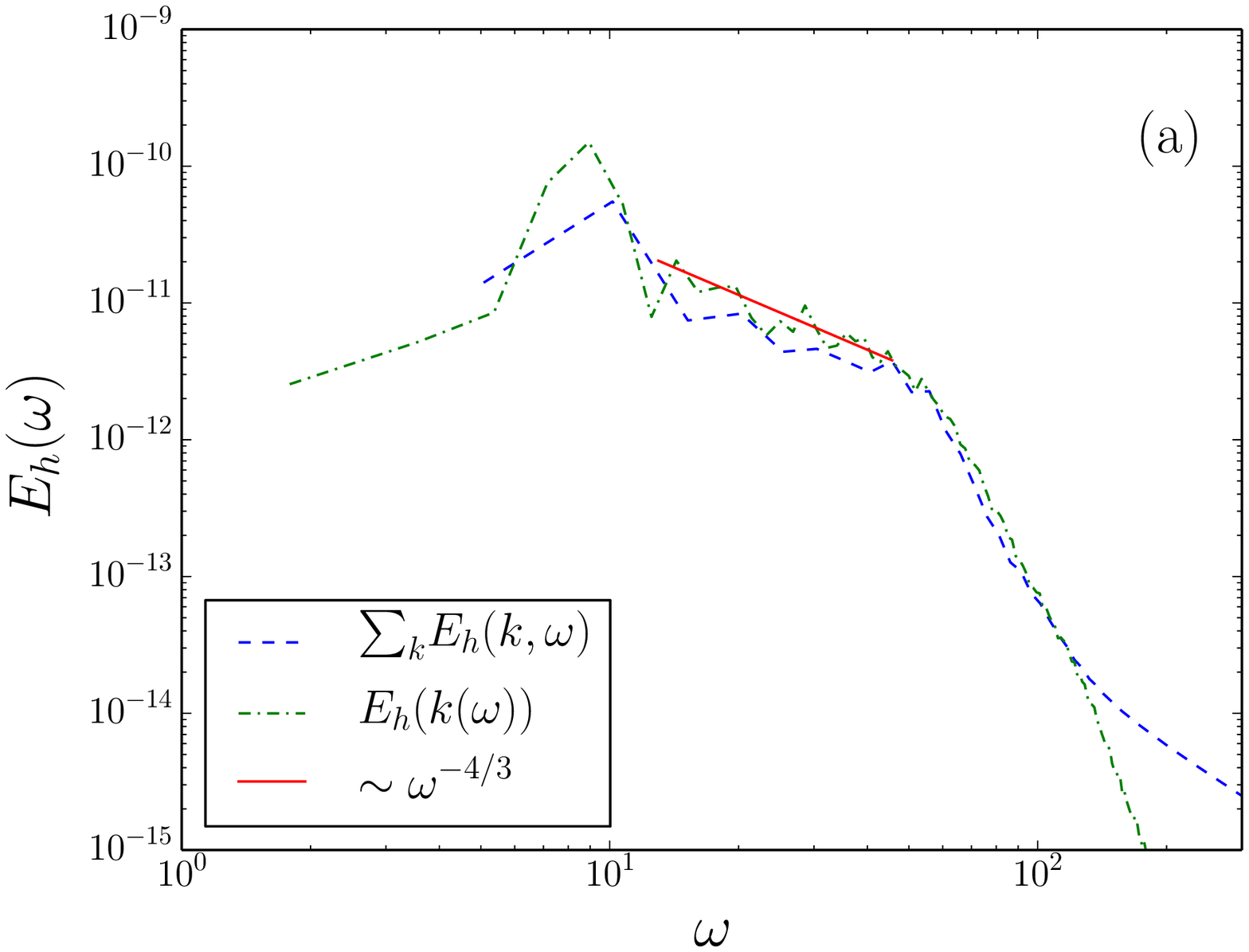}
    \includegraphics[width=0.48\textwidth]{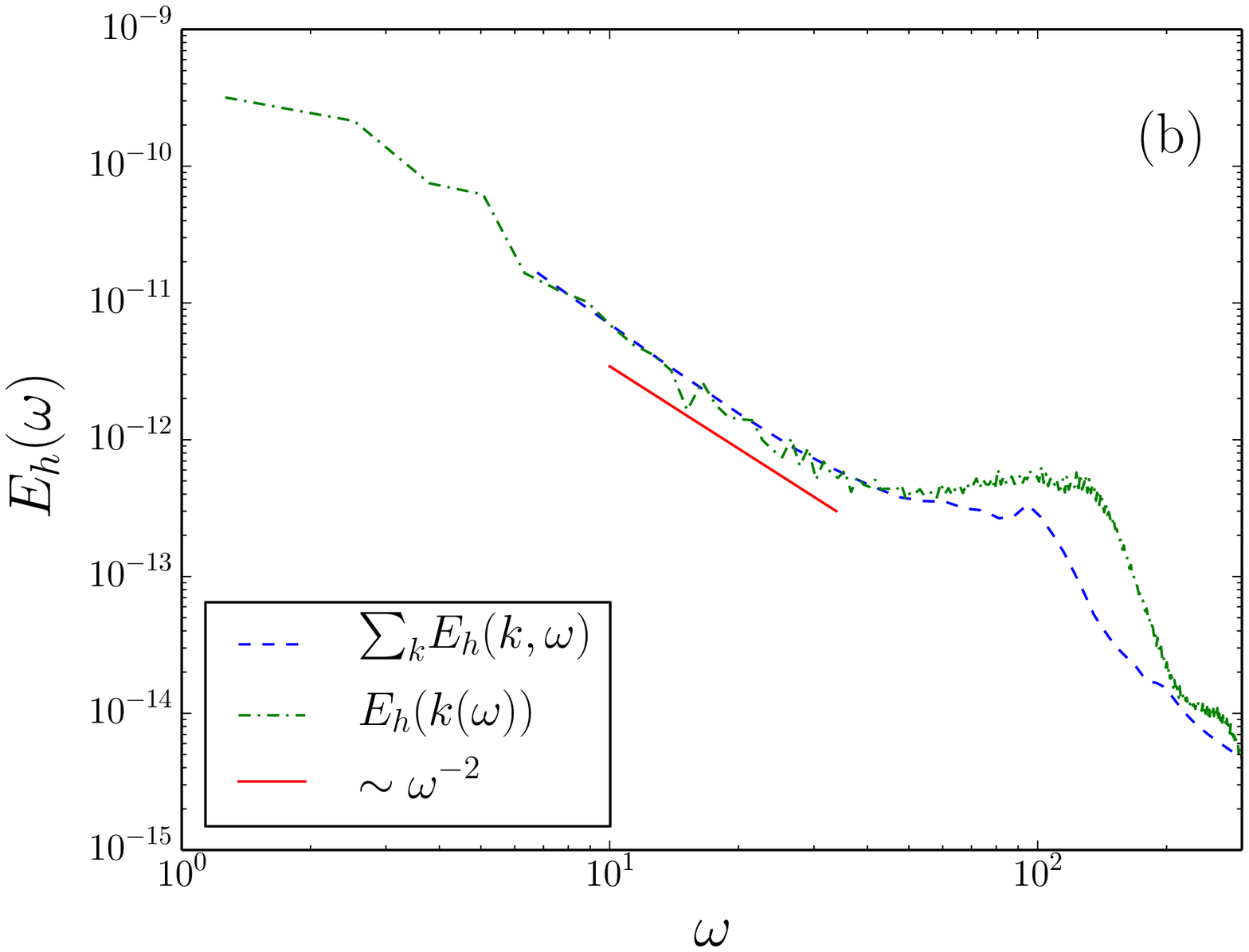}
    \caption{{\it (Color online)} Power spectrum of $h$ as a function of
        the frequency for simulations (a) $A02$ and (b) $B04$. In
        both cases, the spectrum was calculated by summing over all wavenumbers
        in the time and space resolved spectrum, $\textstyle{\sum_k} E_h
        (k,\omega)$, and also by using the dispersion relation given by Eq.~\eqref{rel_disp_bous}
        to estimate the frequency spectrum from the wavenumber spectrum $E_h(k)$. As a reference,
        power laws $\sim \omega^{-4/3}$ and $\sim \omega^{-2}$ are shown in each
        case. The behavior is in good agreement with the one found for $E_h(k)$.}
    \label{edomega}
\end{figure}

As already mentioned, in experiments and simulations $E_h(\omega)$ is
sometimes estimated instead from $E_h(k)$ by using the dispersion relation in the form
$k = k(\omega)$. Figure \ref{edomega} shows the power spectrum of $h$ as a
function of $\omega$ for simulations $A02$ and $B04$. In both cases, the
spectrum was calculated explicitly using Eq.~\eqref{freqspectra}, and
also estimated using the dispersion relation. For each run, the two spectra show a
very good agreement, which can be expected as most of the energy is in
the waves. The behavior of the inertial range in each run is also in good
agreement with the one found previously for $E_h(k)$ in Sec.~\ref{spectra}.

\subsection{Probability density functions} 

\begin{figure}
    \centering
    \includegraphics[width=0.48\textwidth]{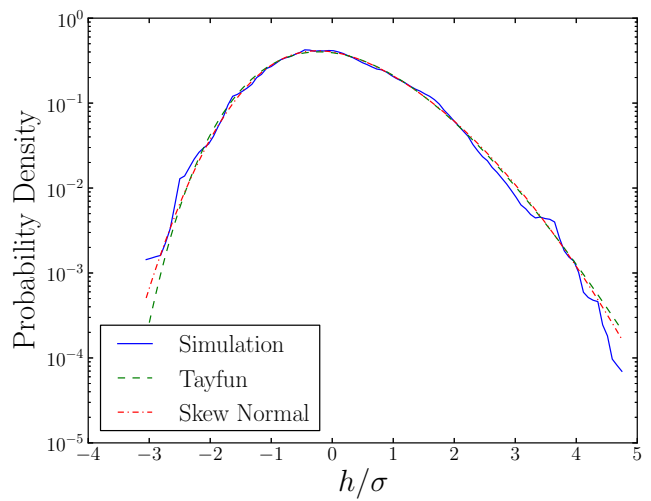}
    \caption{{\it (Color online)} Probability density function of the values of
        $h$ in simulation $A06$ (solid blue line). The dash-dotted (red) line
        indicates a maximum likelihood fit using a skewed normal distribution,
        while the dashed (green) line corresponds to a maximum likelihood fit
        for the Tayfun distribution.}
   \label{distro}
\end{figure}

We calculated the probability density function  (PDF) of the free surface height
for different simulations. Figure \ref{distro} shows the PDF of $h/\sigma$ for
run $A06$, where $\sigma$ is the standard deviation of the surface height. The
probability distribution is asymmetric, with a larger probability of measuring
large values of $h$ than of small values. The shape can be adjusted by two
distributions: We consider a skewed normal distribution 
\cite{azzalini_class_1985},
\begin{equation}
    \label{skew_normal}
    f(x) = \frac{2}{\kappa}\phi\left(\frac{x-\xi}{\kappa}\right)
        \Phi\left(\alpha\frac{x-\xi}{\kappa}\right),   
\end{equation}
where $\kappa$ is the so-called scale parameter (associated with 
the variance of the distribution), $\xi$ is the location parameter
(associated with the mean value), $\alpha$ is the shape parameter
(associated with the skewness), and
\begin{align}
\phi(x) &=\frac{1}{\sqrt{2\pi}}e^{-x^2/2}, \\
\Phi(x) &= \int_{-\infty}^{x} \phi(t)\ \d t = \frac{1}{2} \left[ 1 +
    \operatorname{erf} \left(\frac{x}{\sqrt{2}}\right)\right]. 
\end{align}
We also consider a Tayfun distribution 
\begin{equation}
    p(x) = \int^\infty_0 \frac{e^{-[y^2 + (1-c)^2]/(2 s^2)}}{\pi s c} \d y,
\end{equation}
with $c= \sqrt{1 + 2 s x + y^2}$ and where $s$ is the mean steepness 
of the waves\cite{tayfun_narrow-band_1980}.

For run $A06$, and from a Maximum Likelihood Estimation method for the skewed
normal distribution, the location parameter is $\xi \approx -1.00$, the scale
parameter is $k \approx 1.43$, and the shape parameter is  $\alpha \approx
1.94$. For the same run, and for the Tayfun distribution, the mean steepness of
the waves is $s \approx 0.15$. This latter value is more relevant as the Tayfun
distribution is often used in oceanography and in experiments of surface waves.
In this context, it is interesting to point out that experiments in
\cite{falcon_observation_2011} found similar values for $s$.

This behavior (a PDF of $h$ described correctly by both a skewed normal
distribution and a Tayfun distribution with asymmetry to the left) was observed
in all simulations, no matter what set they belonged to.

\section{Conclusions} 
\label{conclusion}

We studied wave turbulence in shallow water flows in numerical simulations using
the shallow water and Boussinesq models. The equations were solved using grids
up to $2048^2$ points, and the parameters were varied to study different
regimes, including regimes with larger and smaller Reynolds number, and larger
and smaller dispersion, while keeping the Froude number approximately
the same. We summarize below the main conclusions following the same
ordering as in the introduction:

(a) As in previous experimental and observational studies
\cite{smith_equilibrium_2003,kaihatu_asymptotic_2007}, we found that the flows
can be classified in different sets depending on the value of the Reynolds
number (i.e., on the strength of the nonlinearities) and on the level of
dispersion (associated with the fluid depth). A first set ($A$) has smaller
Reynolds numbers and stronger dispersion, a second set ($B$) has larger Reynolds
numbers and weaker or negligible dispersion, and a third set of runs seems to be
transitional between the two.

(b) Runs in sets $A$ and $B$ have different power spectra of the surface height.
Runs in set $A$, with stronger dispersion, present a spectrum compatible (within
statistical uncertainties) with $E_h(k) \sim k^{-4/3}$. This is the spectrum
predicted by weak turbulence theory for the Boussinesq equations
\cite{onorato_four-wave_2008}. Runs in set $B$ with negligible or zero
dispersion (i.e., for a shallower flow) show a spectrum compatible within error
bars with $E_h(k) \sim k^{-2}$. This spectrum can be obtained from
phenomenological arguments coming from strong turbulence
\cite{phillips_equilibrium_1958}. The runs in set $C$ have no discernible
inertial range.

(c) The Boussinesq (dispersive) model tends to develop more power in waves with
short wavelengths than the shallow water model.  This is associated with the
development of a bottleneck for large wavenumbers in the energy spectrum. 

(d) Inspection of the wave and frequency spectrum $E_h(k,\omega)$ 
confirms that most of the energy is in the waves in all the
simulations. In runs in set $B$, most of the energy is concentrated 
in the vicinity the linear dispersion relation for shallow water
waves, which are non-dispersive. In runs in set $A$, the resulting 
non-linear dispersion relation obtained from $E_h(k,\omega)$ has 
two branches: one that corresponds to non-dispersive waves, and 
another corresponding to dispersive waves. The two branches can 
be explained as the result of the superposition of rapidly varying 
waves which ride over slowly varying waves, the latter with 
sufficient amplitude to change whether the former see a shallower 
or deeper fluid.

(e) Independently of the differences between the runs, the probability 
distribution functions of $h$ for the runs in all sets is asymmetric, 
with larger probabilities of finding larger values of $h$ than smaller 
values. The probability distribution functions can be approximated by 
both a skewed normal distribution and a Tayfun distribution 
\cite{tayfun_narrow-band_1980}. In the latter case, the only parameter
of the distribution, the mean steepness of the waves, has values
compatible with those found in observations and experimental studies 
(see \cite{falcon_observation_2011}). The obtained probability density
functions also indicate limitations in the hypothesis of Gaussianity 
of the fields assumed in early theories of weak turbulence. However, 
extensions of the theory to allow for non-Gaussian distributions 
exist and can be found for example in \cite{choi_probability_2004} 
and \cite{lvov_noisy_2004,choi_joint_2005}.

All the results presented here were obtained solving numerically 
equations that do not assume that the flow is inviscid or
irrotational, and with realistic terms for the viscous dissipation. We 
believe this approach can be useful to compare with experiments, as in
experiments vorticity can develop in the flow, and viscosity cannot be 
neglected.

\begin{acknowledgments}
The authors would like to thank Prof. Oliver Buhler and the anonymous referees
for their useful comments. The
authors acknowledge support from grants No. PIP 11220090100825, UBACYT
20020110200359, and PICT 2011-1529 and 2011-1626. PDM and PJC acknowledge
support from the Carrera del Investigador Cient\'ifico of CONICET, and PCdL
acknowledges support from CONICET.
\end{acknowledgments}

\bibliography{Tesis}

\end{document}